\documentclass[useAMS,usegraphicx,usenatbib]{mn2e}

\def\aap     {A\&A}

\def\apj     {ApJ}

\def\araa    {ARA\&A}

\def\mnras   {MNRAS}

\def\pasp    {PASP}
\def\todo{{\Huge $\star$}}

\title{The Prevalence of Cooling Cores in Clusters of Galaxies at $z\approx$~0.15--0.4}
\author[F.~E. Bauer et al.]
{F.~E.~Bauer$^{1}$\thanks{E-mail: feb@ast.cam.ac.uk},
A.~C.~Fabian$^{1}$, 
J.~S.~Sanders$^{1}$, 
S.~W.~Allen$^{1,2}$, 
and R.~M.~Johnstone$^{1}$\\
\footnotesize
$^{1}$Institute of Astronomy, Madingley Road, Cambridge CB3 0HA\\
$^{2}$Kavli Institute for Particle Astrophysics and Cosmology, Stanford
University, 382 Via Pueblo Mall, Stanford, CA 94305-4060, USA.\\
}
\begin{document}
\maketitle

\begin{abstract}
  We present a {\it Chandra} study of 38 X-ray luminous clusters of
  galaxies in the {\it ROSAT} Brightest Cluster Sample (BCS) that lie
  at moderate redshifts ($z\approx0.15$--0.4). Based primarily on
  power ratios and temperature maps, we find that the majority of
  clusters at moderate redshift generally have smooth, relaxed
  morphologies with some evidence for mild substructure perhaps
  indicative of recent minor merger activity. Using spatially-resolved
  spectral analyses, we find that cool cores appear to still be common
  at moderate redshift. At a radius of 50~kpc, we find that at least
  55 per cent of the clusters in our sample exhibit signs of mild
  cooling ($t_{\rm cool} < 10$~Gyr), while in the central bin at least
  34 per cent demonstrate signs of strong cooling ($t_{\rm cool} <
  2$~Gyr). These percentages are nearly identical to those found for
  luminous, low-redshift clusters of galaxies, indicating that there
  appears to be little evolution in cluster cores since $z\approx0.4$
  and suggests that heating and cooling mechanisms may already have
  stabilised by this epoch. Comparing the central cooling times to
  catalogues of central H$\alpha$ emission in BCS clusters, we find a
  strong correspondence between the detection of H$\alpha$ and central
  cooling time. We also confirm a strong correlation between the
  central cooling time and cluster power ratios, indicating that crude
  morphological measures can be used as a proxy for more rigorous
  analysis in the face of limited signal-to-noise data. Finally, we
  find that the central temperatures for our sample typically drop by
  no more than a factor of $\sim$3--4 from the peak cluster
  temperatures, similar to those of many nearby clusters.
\end{abstract}


\section{Introduction}\label{sec:intro}

By the time a cluster of galaxies reaches a relatively relaxed state,
the hot gas in the centre is likely to have a radiative cooling time
which is shorter than the expected cluster age and a temperature which
drops toward the centre, due to energy losses from X-ray emission
\citep[e.g.,][]{Fabian1994}. This effect, termed a ``cooling flow'',
provides a mechanism by which matter can condense out of the hot
intracluster medium (ICM) and is observationally detected as an
enhanced X-ray surface brightness peak in the cores of clusters. The
final state of this cooling gas is still a matter of intense debate
\citep[e.g.,][]{Peterson2001, Tamura2001, Peterson2003, Kaastra2004},
as the temperatures of the cooling gas are found to generally drop by
less than a factor of $\sim$3--4 with very little cooler gas
(presumably due to some form of heating, such as from an active
galactic nucleus).  Despite this cooling ``floor'', and the fact that
the mass of gas cooling to zero represents only a tiny fraction of the
overall hot phase of the ICM, cooling cores remain an observational
phenomenon which inexplicably trace the physical processes by which
clusters of galaxies form. In general, the central regions of a
cluster of galaxies must remain relatively relaxed both to establish
and maintain a cooling core.  Strong disturbances to the cluster, such
as might be caused by merging for example, should mix the cluster gas
and could disrupt the cooling core.  The extent to which the cooling
core can be suppressed or destroyed, in fact, depends on a number of
factors such as the initial strength of the cooling core, how
off-centre the merger is, and the mass ratio of merging systems
\citep[e.g.,][]{McGlynn1984, Edge1992, Allen2001, Gomez2002,
  Ritchie2002}.  Thus, a proper census of cooling cores in clusters as
a function of lookback time should provide important constraints on
the robustness of cooling cores, the nature of cluster buildup and
allow further refinements to structure formation models.

Central surface brightness peaks in clusters of galaxies (the
observational signature of cool cores) appear to be remarkably
widespread, residing in 70--90 per cent of the clusters with $z<0.10$,
and occasionally even dominating their bolometric output
\citep[e.g.,][]{Edge1992, Peres1998}. The ages of these cooling cores,
typically considered to be the time since their last disruption, are
estimated to be $<4$~Gyr in $\sim$50 per cent of the objects
\citep{Allen1997, Allen2001}. By observing galaxy clusters at
$z\approx0.15$--0.4 (i.e., lookback times of 2--4~Gyr), we can probe
the conditions of these objects during a potentially important
disruption period. The question of how common cooling cores are in the
more distant past remains completely unexplored. We report here on a
large sample of moderate redshift clusters with {\it Chanda}
observations. The sub-arcsecond spatial resolution afforded by {\it
  Chandra} is extremely important in this effort, as it allows the
detection and quantification of cooling cores out to much larger
distances than was previously accessible with past X-ray observatories
such as {\it Einstein} and {\it ROSAT}.

We describe our X-ray sample in $\S$\ref{sec:sample}, analysis of the
{\it Chandra} data in $\S$\ref{sec:data}, and finally, our results in
$\S$\ref{sec:discussion}. Throughout this paper, we adopt
$H_{0}=70$~km~s$^{-1}$~Mpc$^{-1}$, $\Omega_{\rm M}=0.3$, and
$\Omega_{\Lambda}=0.7$. Unless explicitly stated otherwise, quoted
errors are for a $1\sigma$ (68 per cent) confidence level.

\section{Sample Selection}\label{sec:sample}

\begin{figure}
\vspace{-0.1in}
\centerline{
\includegraphics[width=9.0cm]{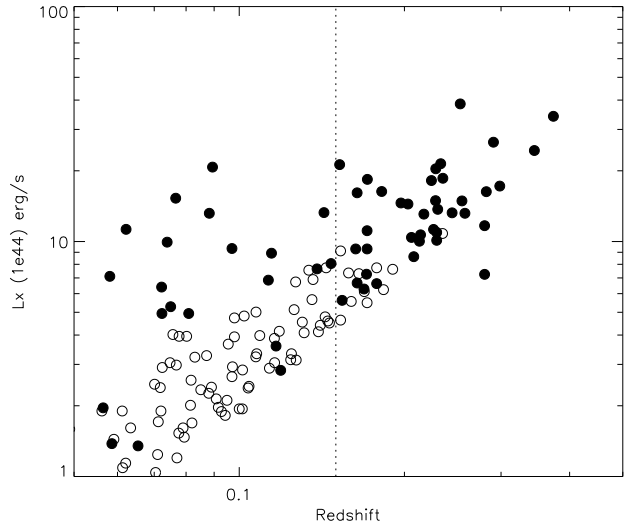}
}
\vspace{-0.1cm} \caption[fig1.ps]{X-ray luminosity versus redshift for
  the BCS sample (circles). Filled circles denote BCS clusters
  observed with {\it Chandra} as of 2003 October. Our sample consists of
  those sources with $z\ge0.15$.\label{fig:select}}
\vspace{-0.3cm}
\end{figure}

We selected our sample from the {\it ROSAT} Brightest Cluster Sample
\citep[BCS;][]{Ebeling1998} and Extended {\it ROSAT} Brightest Cluster
Sample \citep[EBCS;][]{Ebeling2000}, both of which are derived from
ROSAT All-Sky Survey data. When combined (hereafter simply BCS), they
represent one of the largest and most complete X-ray-selected cluster
samples compiled to date. Figure~\ref{fig:select} highlights the
fraction of BCS sources with {\it Chandra} observations, shown in the
0.1--2.4~keV luminosity versus redshift plane. Importantly, a large
fraction of the most luminous and most distant BCS sources have
already been observed by {\it Chandra}. We adopt a selection criterion
of $z>0.15$ to maximize the average redshift of the sample, while
still providing a statistically useful number of objects with {\it
  Chandra} observations. Of the 51 clusters with $z>0.15$, 38 (75 per
cent) have publicly available {\it Chandra} exposures as of 2003
October. The vast majority of these are of sufficient quality (i.e.,
$\ga$5000 counts) to assess crudely the nature of any potential
cooling core. Thus, our sample should be sufficiently complete to
provide reliable statistics on luminous clusters of galaxies at
moderate redshifts. As a local comparison sample, we use the Brightest
55 sample studied by \citet[hereafter B55;][]{Peres1998}.  The B55
sample is a 2--10~keV flux limited sample of X-ray emitting clusters
which is nearly complete and comprised of sources which are all close
enough to have been imaged properly with previous X-ray instruments
(i.e., {\it ROSAT}). To make the B55 sample more comparable with ours,
we have cropped two clusters with $z>0.15$ and excluded clusters with
$L_{0.1--2.4~keV}<4\times 10^{44}$. We have converted the B55 values
to our adopted cosmology.

\begin{table}
\centering
\begin{minipage}{92mm}
\caption{Summary of {\it Chandra} Observations for the $z>0.15$ BCS Sample.}
\label{tab:obs}
\begin{tabular}{lrccrc}
\hline
(1)            & (2)   & (3)         & (4)  & (5)  & (6)\\
Cluster        & ObsID & Date        & ACIS & Exp. & Bgd \\
\hline
A68            & 3250  & 2002-09-07  & I    &   9.9 & 1.039 \\
A115a          & 3233  & 2002-10-07  & I    &  50.0 & 1.045 \\
A115b          & 3233  & 2002-10-07  & I    &  50.0 & 1.045 \\
A267           & 1448  & 1999-10-16  & I    &   7.8 & 1.052 \\
A520           &  528  & 2000-10-10  & I    &   9.3 & 1.010 \\
A586           &  530  & 2000-09-05  & I    &   9.8 & 1.148 \\
A665           & 3586  & 2002-12-28  & I    &  30.1 & 1.068 \\
               &  531  & 1999-12-29  & I    &   8.8 & 1.126 \\
A697           & 4217  & 2002-12-15  & I    &  19.7 & 0.955 \\
               &  532  & 1999-10-21  & I    &   4.4 & 1.765{\rlap *}\\
A750           &  924  & 2000-10-02  & I    &  29.1 & 1.123 \\
A773           &  533  & 2000-09-05  & I    &  11.1 & 1.071 \\
A781           &  534  & 2000-10-03  & I    &   9.9 & 1.032 \\
A963           &  903  & 2000-10-11  & S    &  36.3 & 0.991 \\
A1204          & 2205  & 2001-06-01  & I    &  23.9 & 0.984 \\
A1423          &  538  & 2000-07-07  & I    &   9.5 & 1.019 \\
A1682          & 3244  & 2002-10-19  & I    &   9.6 & 1.784{\rlap !}\\
A1758a         & 2213  & 2001-08-28  & S    &  58.6 & 1.516{\rlap !}\\
A1763          & 3591  & 2003-08-28  & I    &  19.2 & 0.935 \\
A1835          &  495  & 1999-12-11  & S    &  19.7 & 1.327{\rlap !}\\
               &  496  & 2000-04-29  & S    &  10.6 & 1.323{\rlap *}\\
A1914          &  542  & 1999-11-21  & I    &   7.8 & 1.329{\rlap !}\\
A2111          &  544  & 2000-03-22  & I    &  10.1 & 1.084 \\
A2204          &  499  & 2000-07-29  & S    &  10.1 & 1.679{\rlap !}\\
A2218          & 1666  & 2001-08-30  & I    &  46.7 & 1.010 \\
               & 1454  & 1999-10-19  & I    &  11.4 & 1.245{\rlap *}\\
               &  553  & 1999-10-19  & I    &   5.7 & 1.219{\rlap *}\\
A2219          &  896  & 2000-03-31  & S    &  42.5 & 1.573{\rlap !}\\
A2259          & 3245  & 2002-09-16  & I    &   9.9 & 0.990 \\
A2261          &  550  & 1999-12-11  & I    &   9.1 & 1.310{\rlap !}\\
A2294          & 3246  & 2001-12-24  & I    &   9.8 & 1.106 \\
A2390          &  500  & 2000-10-08  & S    &   9.6 & 1.038 \\
               &  501  & 1999-11-05  & S    &   8.8 & 1.344{\rlap *}\\
Hercules A     & 1625  & 2001-07-25  & S    &  14.8 & 1.167{\rlap !}\\
RXJ0439.0+0520 &  527  & 2000-08-29  & I    &   9.6 & 1.003 \\
RXJ0439.0+0715 & 3583  & 2003-01-04  & I    &  18.7 & 0.948 \\
               & 1449  & 1999-10-16  & I    &   6.0 & 1.132 \\
RXJ1532.9+3021 & 1649  & 2001-08-26  & S    &   9.3 & 1.056 \\
               & 1665  & 2001-09-06  & I    &   9.9 & 1.020 \\
RXJ1720.1+2638 & 4361  & 2002-08-19  & I    &  25.7 & 0.990 \\
               & 3224  & 2002-10-03  & I    &  23.8 & 1.035 \\
               & 1453  & 1999-10-19  & I    &   7.8 & 1.291{\rlap *}\\
RXJ2129.6+0005 &  552  & 2000-10-21  & I    &   9.9 & 0.942 \\
Z1953          & 1659  & 2000-10-22  & I    &  24.9 & 0.971 \\
Z2701          & 3195  & 2001-11-04  & S    &  27.0 & 1.563{\rlap !}\\
Z3146          &  909  & 2000-05-10  & I    &  46.4 & 0.987 \\
Z5247          &  539  & 2000-03-23  & I    &   9.1 & 1.116 \\
Z7160          &  543  & 2000-05-19  & I    &   9.6 & 1.055 \\
\hline
\end{tabular}

\smallskip 
{\bf Col. 1:} Cluster name.
{\bf Col. 2:} {\it Chandra} observation number. All observations
are used for image analysis, while only the first observation is used
for spectral analysis.
{\bf Col. 3:} Observation date.
{\bf Col. 4:} Whether observation was ACIS-I or ACIS-S (i.e., cluster
centre was positioned on chip I3 or S3, respectively).
{\bf Col. 5:} Useful exposure time after the data were cleaned and
background flares removed, in units of ks.
{\bf Col. 6:} Ratio of the average 0.3--10~keV background count rate
for each observation versus the ACIS quiescent-background calibration
values (i.e., 0.31 cts s$^{-1}$ per ACIS-I chip and 0.86 cts s$^{-1}$
per ACIS-S3 chip; see Sect.~\ref{sec:prep}). Symbols indicate that the
observation has a high background, but was used only for image
analysis (``*'') or for both image and spectral analysis (``!'').
\end{minipage}
\end{table}

\begin{table*}
\scriptsize
\centering
\begin{minipage}{174mm}
\caption{General Properties of the Cool Core Sample.}
\label{tab:props1}
\begin{tabular}{lccrcrrccrrrr}
\hline
(1)            & (2)              &   (3)       & (4)   & (5)    & (6)         & (7)         & (8)       & (9)   & (10)           & (11)      & (12)      & (13)       \\
Cluster        & Position (J2000) & $N_{\rm H}$ & $kT$  & $z$    & $F_{\rm X}$ & $L_{\rm X}$ & $H\alpha$ & Mor. & $R_{\rm cent}$ & $t_{\rm cool}$ & $t_{\rm 50}$ & $t_{\rm 180}$ \\
\hline
A68            & 00$^{h}$37$^{m}$06\fs1 $+$09$^{\circ}$09\arcmin34\arcsec & 4.6 &    10.0 & 0.2546 &   5.5   &   14.89  & ? & R &  41.3 & $ 24.0^{+ 10.6}_{- 10.8}$ & $ 24.0^{+ 10.6}_{- 10.8}$ & $ 57.5^{+  0.0}_{- 22.6}$ \\
A115a*         & 00$^{h}$55$^{m}$50\fs9 $+$26$^{\circ}$24\arcmin35\arcsec & 5.2 &     9.8 & 0.1971 &   9.0   &   14.59  & Y & D &  17.4 & $  1.6^{+  0.2}_{-  0.2}$ & $  4.0^{+  0.7}_{-  0.6}$ & $ 23.0^{+  5.8}_{-  4.1}$ \\
A115b*         & 00$^{h}$55$^{m}$59\fs4 $+$26$^{\circ}$20\arcmin13\arcsec & 5.2 &     9.8 & 0.1971 &   9.0   &   14.59  & ? & D &  43.1 & $ 22.0^{+  9.7}_{-  6.2}$ & $ 22.0^{+  9.7}_{-  6.2}$ & $ 34.1^{+ 62.1}_{- 10.4}$ \\
A267           & 01$^{h}$52$^{m}$42\fs1 $+$01$^{\circ}$00\arcmin38\arcsec & 2.8 &     9.7 & 0.2300 &   6.2   &   13.71  & N & R &  36.9 & $ 25.1^{+  4.2}_{- 13.1}$ & $ 25.1^{+  4.2}_{- 13.1}$ & $ 22.0^{+ 20.7}_{-  6.7}$ \\
A520           & 04$^{h}$54$^{m}$09\fs8 $+$02$^{\circ}$55\arcmin16\arcsec & 7.4 &     9.8 & 0.2030 &   8.4   &   14.44  & N & D &  52.7 & $ 14.8^{+ 13.6}_{-  4.8}$ & $ 13.6^{+ 12.5}_{-  4.4}$ & $ 24.5^{+ 23.2}_{-  7.8}$ \\
A586           & 07$^{h}$32$^{m}$20\fs3 $+$31$^{\circ}$37\arcmin56\arcsec & 5.4 &     8.7 & 0.1710 &   9.1   &   11.12  & N & R &  30.2 & $  8.8^{+  4.3}_{-  2.0}$ & $  8.8^{+  4.3}_{-  2.0}$ & $ 15.8^{+  4.3}_{-  2.8}$ \\
A665           & 08$^{h}$30$^{m}$58\fs0 $+$65$^{\circ}$50\arcmin54\arcsec & 4.0 &     8.3 & 0.1818 &  11.8   &   16.33  & N & D &  38.4 & $ 18.2^{+ 14.9}_{-  5.5}$ & $ 18.2^{+ 14.9}_{-  5.5}$ & $ 16.2^{+  4.4}_{-  3.0}$ \\
A697           & 08$^{h}$42$^{m}$57\fs7 $+$36$^{\circ}$21\arcmin57\arcsec & 3.4 &    10.5 & 0.2820 &   5.0   &   16.30  & N & R &  43.6 & $ 28.4^{+  5.2}_{- 12.7}$ & $ 28.4^{+  5.2}_{- 12.7}$ & $ 25.3^{+ 21.5}_{-  8.4}$ \\
A750           & 09$^{h}$09$^{m}$12\fs2 $+$10$^{\circ}$58\arcmin36\arcsec & 3.6 &     8.1 & 0.1630 &   8.3   &    9.30  & N & R &  22.9 & $  7.2^{+  2.5}_{-  1.6}$ & $  7.2^{+  2.5}_{-  1.6}$ & $ 25.1^{+  9.2}_{-  5.5}$ \\
A773           & 09$^{h}$17$^{m}$52\fs9 $+$51$^{\circ}$43\arcmin37\arcsec & 1.5 &     9.4 & 0.2170 &   6.7   &   13.08  & N & R &  33.1 & $ 24.8^{+ 11.6}_{- 15.8}$ & $ 24.8^{+ 11.6}_{- 15.8}$ & $ 63.4^{+  0.0}_{- 24.4}$ \\
A781           & 09$^{h}$20$^{m}$25\fs4 $+$30$^{\circ}$30\arcmin13\arcsec & 1.9 &    10.8 & 0.2984 &   4.7   &   17.22  & ? & 2 &  81.7 & $ 17.4^{+  8.9}_{-  4.6}$ & $  7.2^{+  3.7}_{-  1.9}$ & $ 17.4^{+  8.9}_{-  4.6}$ \\
A963           & 10$^{h}$17$^{m}$03\fs5 $+$39$^{\circ}$02\arcmin54\arcsec & 1.4 &     8.6 & 0.2060 &   5.9   &   10.41  & N & R &  20.8 & $  4.5^{+  0.9}_{-  0.7}$ & $  4.5^{+  0.9}_{-  0.7}$ & $ 23.1^{+  6.0}_{-  4.5}$ \\
A1204          & 11$^{h}$13$^{m}$20\fs4 $+$17$^{\circ}$35\arcmin38\arcsec & 1.4 &     7.5 & 0.1904 &   5.0   &    7.62  & Y & R &   9.5 & $  1.1^{+  0.3}_{-  0.2}$ & $  2.1^{+  0.4}_{-  0.3}$ & $ 15.4^{+  2.2}_{-  1.8}$ \\
A1423          & 11$^{h}$57$^{m}$17\fs2 $+$33$^{\circ}$36\arcmin39\arcsec & 1.6 &     8.5 & 0.2130 &   5.3   &   10.03  & N & R &  24.4 & $  4.2^{+  1.7}_{-  1.1}$ & $  4.2^{+  1.7}_{-  1.1}$ & $ 22.5^{+ 13.0}_{-  6.5}$ \\
A1682          & 13$^{h}$06$^{m}$51\fs8 $+$46$^{\circ}$33\arcmin13\arcsec & 1.1 &     8.9 & 0.2260 &   5.3   &   11.26  & N & 2 &  49.4 & $ 34.6^{+ 35.8}_{- 18.2}$ & $ 34.6^{+ 35.8}_{- 18.2}$ & $113.0^{+  0.0}_{- 48.5}$ \\
A1758a         & 13$^{h}$32$^{m}$43\fs5 $+$50$^{\circ}$32\arcmin44\arcsec & 1.1 &     9.1 & 0.2800 &   3.6   &   11.68  & N & 2 &  62.7 & $ 31.8^{+ 20.5}_{-  9.6}$ & $ 21.4^{+ 13.8}_{-  6.5}$ & $ 38.0^{+ 27.0}_{- 11.6}$ \\
A1763          & 13$^{h}$35$^{m}$18\fs7 $+$40$^{\circ}$59\arcmin59\arcsec & 0.9 &    10.0 & 0.2279 &   6.9   &   14.93  & N & R &  45.6 & $ 13.6^{+  8.9}_{-  3.8}$ & $ 13.6^{+  8.9}_{-  3.8}$ & $ 36.6^{+ 33.4}_{- 10.0}$ \\
A1835          & 14$^{h}$01$^{m}$02\fs0 $+$02$^{\circ}$52\arcmin40\arcsec & 2.2 &    14.8 & 0.2528 &  14.7   &   38.53  & Y & R &  11.5 & $  0.6^{+  0.1}_{-  0.1}$ & $  1.0^{+  0.1}_{-  0.1}$ & $ 12.5^{+  4.4}_{-  2.4}$ \\
A1914          & 14$^{h}$26$^{m}$00\fs8 $+$37$^{\circ}$49\arcmin37\arcsec & 1.0 &    10.8 & 0.1712 &  15.0   &   18.39  & N & 2 &  29.6 & $ 14.2^{+ 11.5}_{-  6.5}$ & $ 14.2^{+ 11.5}_{-  6.5}$ & $ 35.9^{+  0.0}_{- 12.4}$ \\
A2111          & 15$^{h}$39$^{m}$41\fs1 $+$34$^{\circ}$25\arcmin10\arcsec & 2.0 &     8.8 & 0.2290 &   5.0   &   10.94  & N & R &  50.1 & $ 21.4^{+ 38.1}_{-  9.5}$ & $ 21.4^{+ 38.0}_{-  9.5}$ & $ 38.7^{+ 40.8}_{- 19.8}$ \\
A2204          & 16$^{h}$32$^{m}$47\fs0 $+$05$^{\circ}$34\arcmin32\arcsec & 5.3 &    11.4 & 0.1524 &  21.9   &   21.25  & Y & R &   7.7 & $  0.3^{+  0.0}_{-  0.0}$ & $  1.4^{+  0.2}_{-  0.1}$ & $ 17.9^{+  4.6}_{-  3.2}$ \\
A2218          & 16$^{h}$35$^{m}$52\fs0 $+$66$^{\circ}$12\arcmin34\arcsec & 3.4 &     6.7 & 0.1710 &   7.5   &    9.30  & ? & R &  34.0 & $ 30.9^{+ 10.7}_{- 12.0}$ & $ 30.9^{+ 10.7}_{- 12.0}$ & $ 18.4^{+  3.1}_{-  3.1}$ \\
A2219          & 16$^{h}$40$^{m}$20\fs2 $+$46$^{\circ}$42\arcmin30\arcsec & 1.7 &    11.4 & 0.2281 &   9.5   &   20.40  & N & R &  41.1 & $ 29.1^{+  0.0}_{-  4.5}$ & $ 29.1^{+  0.0}_{-  4.5}$ & $ 24.9^{+ 10.3}_{-  5.6}$ \\
A2259          & 17$^{h}$20$^{m}$08\fs4 $+$27$^{\circ}$40\arcmin10\arcsec & 3.8 &     7.1 & 0.1640 &   5.9   &    6.66  & N & R &  28.9 & $  9.0^{+  6.2}_{-  2.8}$ & $  9.0^{+  6.2}_{-  2.8}$ & $ 15.3^{+  8.7}_{-  3.5}$ \\
A2261          & 17$^{h}$22$^{m}$27\fs0 $+$32$^{\circ}$07\arcmin57\arcsec & 3.2 &    10.8 & 0.2240 &   8.7   &   18.18  & N & R &  18.6 & $  3.0^{+  1.7}_{-  0.9}$ & $  5.0^{+  2.9}_{-  1.4}$ & $ 22.3^{+ 23.5}_{-  8.4}$ \\
A2294          & 17$^{h}$24$^{m}$05\fs5 $+$85$^{\circ}$53\arcmin15\arcsec & 6.1 &     7.1 & 0.1780 &   5.0   &    6.62  & Y & R &  31.8 & $ 13.1^{+ 15.5}_{-  4.5}$ & $ 13.1^{+ 15.5}_{-  4.5}$ & $ 64.5^{+  0.0}_{- 17.0}$ \\
A2390          & 21$^{h}$53$^{m}$36\fs9 $+$17$^{\circ}$41\arcmin46\arcsec & 6.6 &    11.6 & 0.2329 &   9.6   &   21.44  & Y & R &  20.8 & $  1.9^{+  0.4}_{-  0.3}$ & $  1.9^{+  0.4}_{-  0.3}$ & $ 11.7^{+  3.4}_{-  2.4}$ \\
Hercules A     & 16$^{h}$51$^{m}$08\fs2 $+$04$^{\circ}$59\arcmin33\arcsec & 6.0 &     6.6 & 0.1540 &   5.6   &    5.62  & ? & R &  17.1 & $  1.3^{+  0.1}_{-  0.1}$ & $  1.3^{+  0.1}_{-  0.1}$ & $ 22.4^{+  6.0}_{-  4.5}$ \\
RXJ0439.0+0520 & 04$^{h}$39$^{m}$02\fs4 $+$05$^{\circ}$20\arcmin43\arcsec & 9.7 &     7.9 & 0.2080 &   4.8   &    8.62  & Y & R &   7.8 & $  0.5^{+  0.1}_{-  0.1}$ & $  4.1^{+  2.2}_{-  1.1}$ & $ 19.0^{+  7.5}_{-  4.9}$ \\
RXJ0439.0+0715 & 04$^{h}$39$^{m}$00\fs5 $+$07$^{\circ}$16\arcmin00\arcsec & 1.0 &     9.5 & 0.2443 &   5.3   &   13.25  & N & R &  30.6 & $  6.1^{+  1.5}_{-  1.0}$ & $  6.1^{+  1.5}_{-  1.0}$ & $ 20.0^{+  9.3}_{-  5.1}$ \\
RXJ1532.9+3021 & 15$^{h}$32$^{m}$53\fs8 $+$30$^{\circ}$20\arcmin59\arcsec & 2.1 &    12.5 & 0.3615{\rlap !} &   5.0   &   24.40  & Y & R &  10.1 & $  0.5^{+  0.1}_{-  0.1}$ & $  0.7^{+  0.3}_{-  0.1}$ & $ 15.2^{+  5.3}_{-  3.8}$ \\
RXJ1720.1+2638 & 17$^{h}$20$^{m}$10\fs1 $+$26$^{\circ}$37\arcmin29\arcsec & 3.9 &    10.2 & 0.1640 &  14.3   &   16.12  & Y & R &  13.6 & $  1.9^{+  0.4}_{-  0.3}$ & $  1.9^{+  0.3}_{-  0.2}$ & $ 15.2^{+  2.6}_{-  1.9}$ \\
RXJ2129.6+0005 & 21$^{h}$29$^{m}$39\fs9 $+$00$^{\circ}$05\arcmin18\arcsec & 4.3 &    11.0 & 0.2350 &   8.1   &   18.59  & Y & R &  13.1 & $  0.7^{+  0.1}_{-  0.1}$ & $  3.2^{+  1.6}_{-  0.8}$ & $ 20.7^{+ 10.7}_{-  5.7}$ \\
Z1953          & 08$^{h}$50$^{m}$07\fs0 $+$36$^{\circ}$04\arcmin20\arcsec & 3.1 &    14.5 & 0.3737 &   6.1   &   34.12  & ? & R &  33.3 & $ 28.8^{+  4.6}_{- 17.4}$ & $ 28.8^{+  4.6}_{- 17.4}$ & $ 23.4^{+ 25.7}_{-  9.2}$ \\
Z2701          & 09$^{h}$52$^{m}$49\fs2 $+$51$^{\circ}$53\arcmin05\arcsec & 0.9 &     8.7 & 0.2140 &   5.6   &   10.68  & Y & R &  13.4 & $  1.2^{+  0.3}_{-  0.1}$ & $  2.9^{+  0.6}_{-  0.4}$ & $ 19.2^{+  4.3}_{-  3.1}$ \\
Z3146          & 10$^{h}$23$^{m}$39\fs6 $+$04$^{\circ}$11\arcmin12\arcsec & 2.7 &    12.8 & 0.2906 &   7.7   &   26.47  & Y & R &  11.4 & $  0.6^{+  0.0}_{-  0.0}$ & $  1.5^{+  0.2}_{-  0.2}$ & $ 12.0^{+  1.9}_{-  1.6}$ \\
Z5247          & 12$^{h}$34$^{m}$21\fs8 $+$09$^{\circ}$47\arcmin03\arcsec & 1.7 &     8.5 & 0.2290 &   4.6   &   10.12  & N & 2 &  95.5 & $169.0^{+ 43.3}_{-115.4}$ & $ 57.3^{+ 14.7}_{- 39.1}$ & $169.0^{+ 43.3}_{-115.4}$ \\
Z7160          & 14$^{h}$57$^{m}$15\fs0 $+$22$^{\circ}$20\arcmin34\arcsec & 3.1 &     9.6 & 0.2578 &   4.8   &   13.19  & Y & R &   9.3 & $  0.7^{+  0.3}_{-  0.2}$ & $  1.5^{+  0.5}_{-  0.3}$ & $ 21.2^{+ 11.8}_{-  5.5}$ \\
\hline
\end{tabular}

\smallskip 
{\bf Col. 1:} Cluster name. *The {\it ROSAT} derived temperature, flux
and luminosity given for A0115 were measured for the source as a
whole, rather than as two separate, merging clusters; given their
projected separation, we treat them as individual clusters in our
analysis.
{\bf Col. 2:} Cluster J2000 centroid position as determined from the
{\it Chandra} image after excising contaminating point sources.
{\bf Col. 3:} Galactic absorption column, in units of
$10^{20}$~cm$^{-2}$.
{\bf Col. 4:} BCS temperature as derived from $L_{\rm X}$--$T$ relation,
in units of keV.
{\bf Col. 5:} BCS redshift.
{\bf Col. 6:} BCS unabsorbed X-ray flux in the 0.1--2.4 keV band, in units
of $10^{-12}$~erg$^{-1}$~s$^{-1}$~cm$^{-2}$ 
{\bf Col. 7:} BCS intrinsic X-ray luminosity in the 0.1-2.4 keV band
(cluster rest frame), in units of $10^{44}$~erg$^{-1}$~s.
Values in columns 3--7 are taken from \citet{Ebeling1998}, except for
the redshift of RXJ1532.9$+$3021 denoted by ``!'', which was
determined more accurately by \citet{Crawford1999}.
{\bf Col. 8:} Detection of $H\alpha$ line emission as determined by 
\citet{Crawford1999}. Y$=$yes, N$=$no, ?$=$not observed.
{\bf Col. 9:} Cluster morphology. Assessed by eye.
R$=$regular, D$=$Disturbed, 2$=$Double peaked, often disturbed.
{\bf Col. 10:} Central cooling bin radius at which the central cooling
time is measured, in units of kpc.
{\bf Col. 11:} Central cooling time, in units of Gyr.
{\bf Col. 12:} Cooling time at 50~kpc, in units of Gyr. Taken from the
closest bin below 50~kpc, or interpolated if the central bin is above
50~kpc.
(see $\S$~\ref{sec:cool_profs} for details).
{\bf Col. 13:} Cooling time at 180~kpc, in units of Gyr. Taken from the
closest bin below 180~kpc (see $\S$~\ref{sec:cool_profs} for details).
\end{minipage}
\end{table*}
 
\section{Data}\label{sec:data}

\subsection{Preparation}\label{sec:prep}

The {\it Chandra} data (see Table~\ref{tab:obs}) were processed and
cleaned using the CIAO software and calibration files (CIAO v3.0.2,
CALDB 2.27) provided by the {\it Chandra} X-ray Center (CXC), but also
with FTOOLS (v5.3.1) and custom software. We began by reprocessing the
data with the CIAO tool ACIS\_PROCESS\_EVENTS to remove the standard
pixel randomisation and to flag potential ACIS background events for
data observed in Very Faint (VF) mode. The data were then corrected
for the radiation damage sustained by the CCDs during the first few
months of {\it Chandra} operations using the charge transfer
inefficiency correction procedure of
\citet{Townsley2002}.\footnote{For details see
  http://www.astro.psu.edu/users/townsley/cti/.}  Following
CTI-correction, we performed standard grade selection,
excluding all bad columns, bad pixels, cosmic-ray
afterglows, and VF-mode background events when applicable.  We used
only data taken during times within the CXC-generated good-time
intervals. Furthermore, light curves for each dataset were made to
screen for periods of high background; periods exceeding 3$\sigma$ of
the mean rate were removed. 

Backgrounds rates were then estimated from local regions as free of
cluster emission as possible; in a few instances where large clusters
were observed with ACIS-S, our background rate estimates are likely to
be contaminated upward by faint cluster emission. Ratios of the
average 0.3--10~keV background count rate for each observation to that
of ACIS quiescent-background calibration values (i.e., 0.31 cts
s$^{-1}$ per ACIS-I chip and 0.86 cts s$^{-1}$ per ACIS-S3 chip) are
listed in Table~\ref{tab:obs}.  The vast majority of the observations
have average background rates consistent to better than 15 per cent
with ACIS quiescent-background calibration measurements.\footnote{See
  http://cxc.harvard.edu/contrib/maxim/bg/index.html.}

For such observations, the blank sky backgrounds of M.~Markevitch,
processed in an identical manner to the original sample datasets, were
used to characterise the X-ray background.\footnote{See
  http://hea-www.harvard.edu/$\sim$maxim/axaf/acisbg/.}  For the
remaining observations with high background rates (noted in column 7
of Table~\ref{tab:obs} by ``!'' or ``*''; the symbols respectively
denote observations used for both imaging and spectral analysis and
imaging analysis only), local backgrounds were determined from regions
as free of cluster emission as possible. The sample clusters are all
small enough that we easily can extract source free background regions
on the I0-I3 chips for ACIS-I. However, we caution that background
regions on the S3 chip for ACIS-S may still include some faint cluster
emission, thereby affecting the spectral analysis for a few of the
outermost regions/annuli. Note that the degree of such background
contamination is thus too small to significanly bias any of our
conclusions.

Observations were combined for imaging and source detection purposes,
using the 0.3--7.0~keV band. Contaminating point sources were found
using WAVDETECT and masked out of the event lists using their 95 per
cent encircled-energy radii \citetext{e.g., Feigelson, Broos, \&
  Gaffney 2000; \citealp{Jerius2000}; M.~Karovska and P.~Zhao 2001,
  private communication}.\footnote{Feigelson et~al.  (2000) is
  available at
  http://www.astro.psu.edu/xray/acis/memos/memoindex.html.} Cluster
centres were determined using a simple centroiding method on the
masked 0.3--7.0~keV X-ray images (these are given in
Table~\ref{tab:props1}). Radial profiles for each cluster were fit
initially with single 1-D beta models, followed by double 1-D beta
models if the fit resulted in a $\chi^{2}_{\nu}>2.0$. The core radius,
beta index, and amplitude were allowed to vary over sensible limits.
Postage stamp 0.3--7.0~keV X-ray images and radial profiles for each
cluster are provided in Figure~\ref{fig:images}.

\begin{figure*}
\vspace{-0.1cm} \caption[fig2.ps]{ \todo Images are too large to
include in astro-ph version; download full version at
http://www.astro.columbia.edu/~feb/Bauer.ps.gz
{\it Left:} $8\arcmin\times8\arcmin$
colour images for each BCS cluster in the sample with {\it
Chandra} data, comprised from the 0.5--1.0~keV
(red), 1--2~keV (green), and 2--7~keV (blue) X-ray energy bands. The
images have been spatially binned by a factor of 2, smoothed with a
$\sigma=$3~pixel Gaussian smoothing scale, and logarithmically scaled
\citep[following the methods of][]{Lupton2004}. The white cross
denotes the cluster centroid, while the white circle
indicates a radius of 0.5~Mpc at the distance of each cluster.
{\it Middle:} Radial profiles for each BCS cluster in the
sample.
{\it Right:} Projected temperature maps for each BCS cluster in the
sample, scaled between 0--20~keV. The maps have been cropped to the
0.5~Mpc radius shown in the colour images to speed computation. The
signal-to-noise of the data in each region (which varies between
20--40) is shown in the bottom right corner of each map.
\label{fig:images}}
\vspace{-0.3cm}
\end{figure*} 

From inspection of the postage stamp images and radial profiles in
Figure~\ref{fig:images}, the majority of the clusters appear relatively
regular and relaxed, although a notable minority exhibit clear
asymmetric morphologies suggestive of recent merger activity or sharp
discontinuities indicative of inhomogeneities (e.g., cold fronts or
shocks).  We calculated power ratios for each cluster to quantify
obvious morphological structure within the sample in an objective
manner.

\subsection{Power Ratios}\label{sec:power}

Power ratios allow comparisons of cluster morphologies on a common
scale, effectively normalising cluster fluxes within a specified
aperture.  Following the prescription in \citet{Buote1996}, we
determined power ratios for each of the 38 clusters in the sample.  As
described therein, the power ratios are derived from the multipole
expansion of the two-dimensional gravitational potential owing to
matter interior to a user-defined aperture, essentially measuring the
square of the ratio of higher order multipole moments to the monopole
moment (i.e., $P_{1}/P_{0}$, $P_{2}/P_{0}$, $P_{3}/P_{0}$...).  The
coordinate system is chosen such that $P_{1}$ is trivially zero, and
thus the power ratios $P_{2}/P_{0}$, $P_{3}/P_{0}$, and $P_{4}/P_{0}$
provide the most useful information for elucidating basic
morphological/evolutionary trends \citep[e.g.,]{Buote1996}. Low values
of $P_{m}/P_{0}$ indicate highly relaxed, compact and symmetric
clusters on the scale of the aperture used, while high values indicate
either relatively flat, spread out clusters or clusters with
significant substructure or disturbed morphologies. Odd multipole
ratios yield the largest differences between single and bimodal
clusters, essentially vanishing for single-component clusters.
$P_{3}/P_{0}$ is particularly sensitive to unequal-sized bimodals,
with higher values for stronger bimodal asymmetry. Even multipole
ratios are also able to separate bimodals from low-ellipticity
single-component clusters, but demonstrate more considerable overlap
between flatter single-component clusters and moderately bimodal
clusters than the odd terms.

\begin{figure*}
\vspace{-0.1in}
\centerline{
\includegraphics[width=6.0cm]{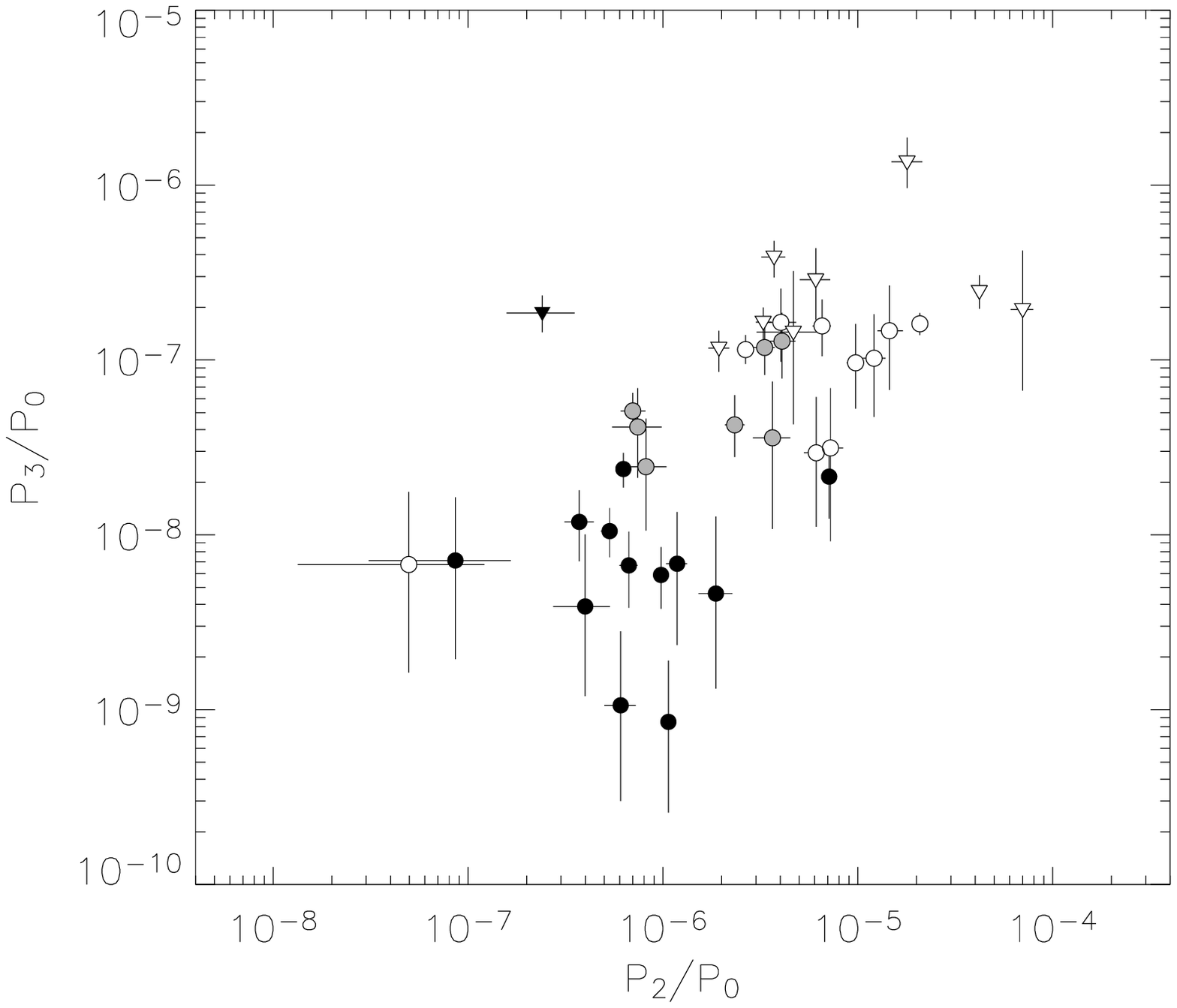}
\hglue-0.0in{\includegraphics[width=6.0cm]{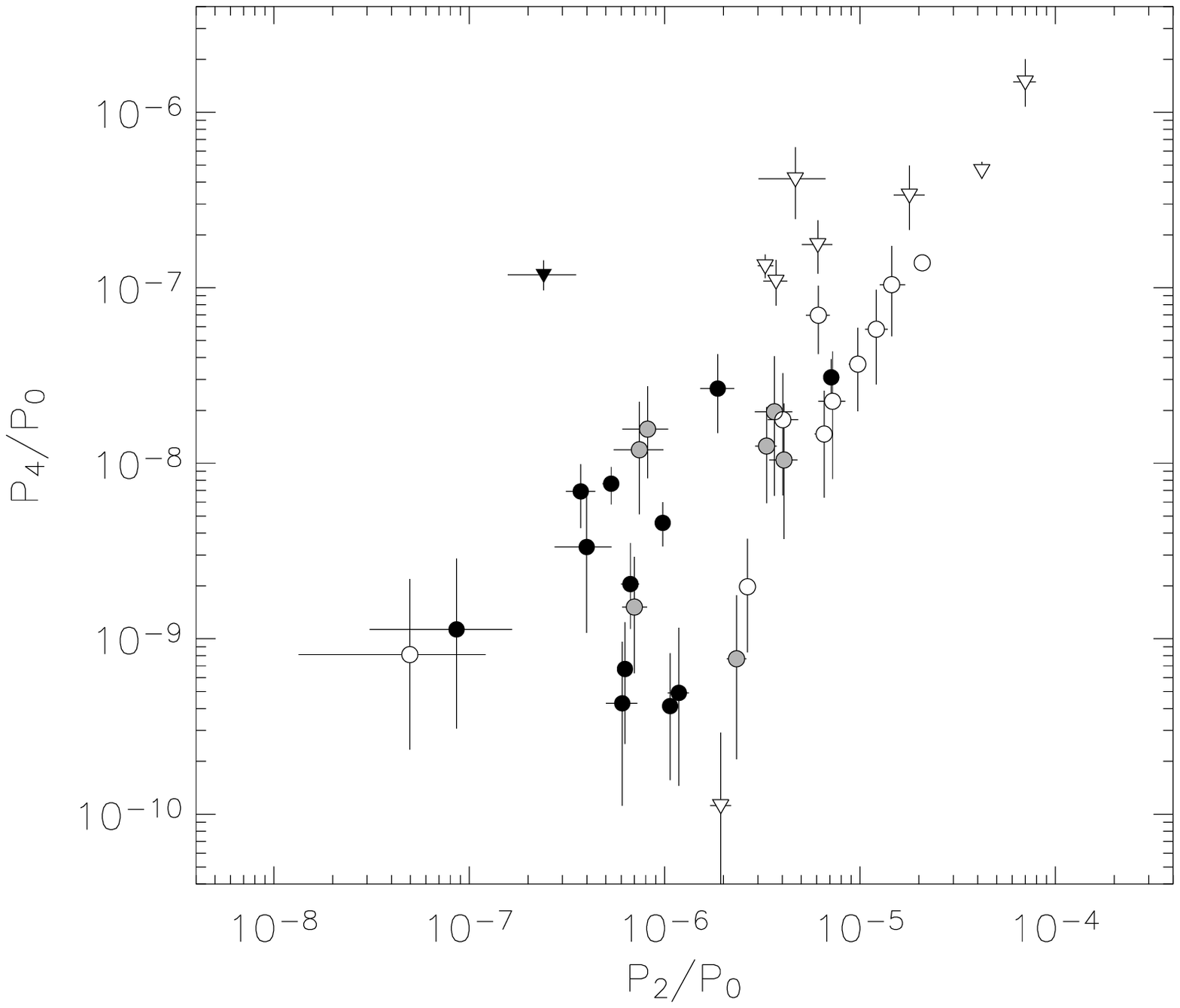}}
\hglue-0.0in{\includegraphics[width=6.0cm]{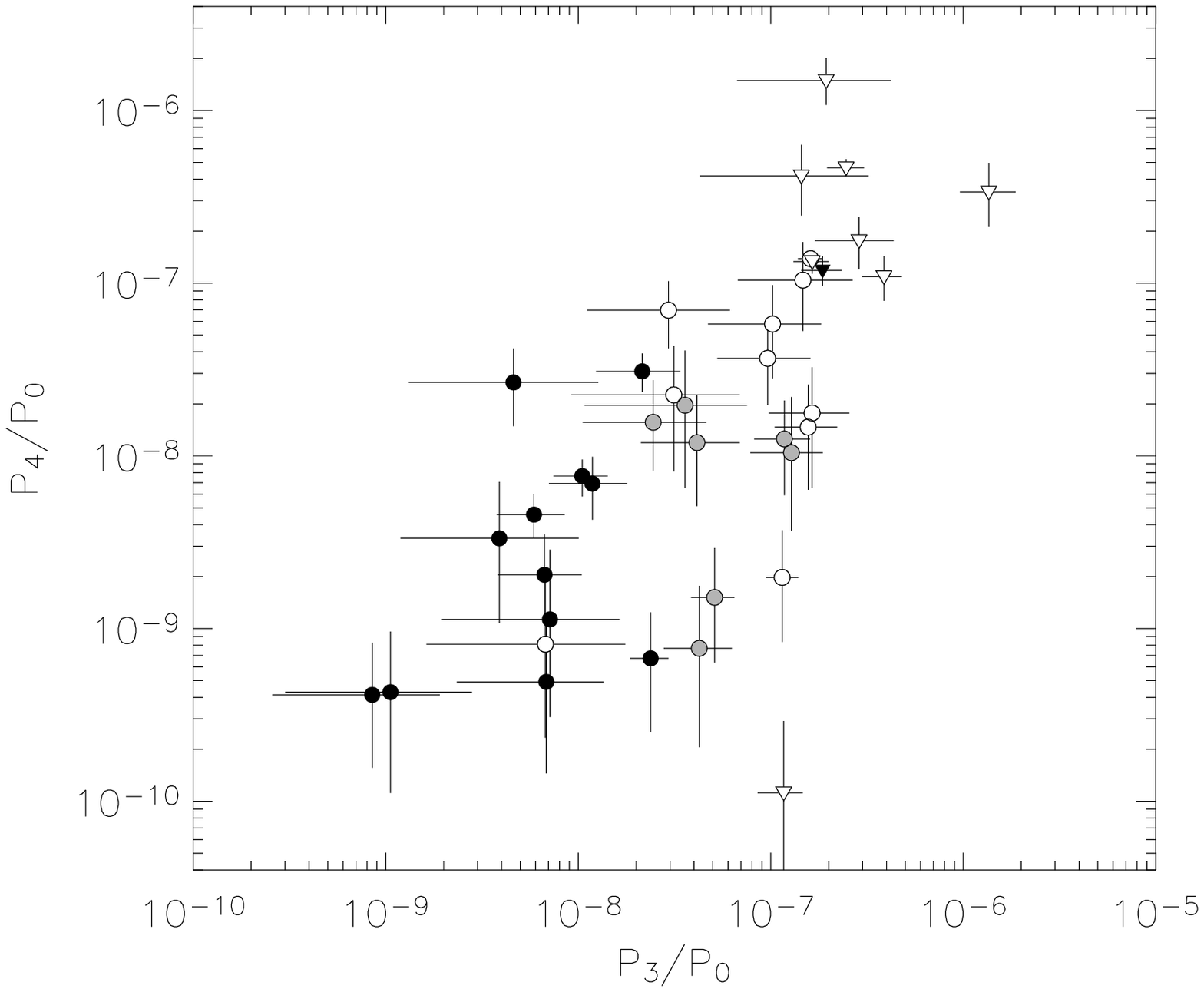}}
}
\vspace{-0.1cm} \caption[fig3.ps]{Power ratios for the 38 BCS clusters 
  in the sample using a 0.5~Mpc aperture. Filled  black, grey, and empty
  symbols denote clusters with strong cool cores ($t_{c}<2$~Gyr),
  weak cool cores ($t_{c}=2$--10~Gyr), and no obvious cool cores
  ($t_{c}>10$~Gyr), respectively. Circles denote morphologically
  regular clusters, while triangles denote double-peaked or disturbed
  clusters. A2294 (open circle in the lower left of the first two
  panels) is again an exception, being a very compact cluster without
  a cool core.  \label{fig:power2}}
\vspace{-0.3cm}
\end{figure*}

The power ratios for the 38 BCS clusters are plotted in
Figure~\ref{fig:power2}, as well as listed in Table~\ref{tab:props2}.
We used intrinsic physical apertures of 0.5~Mpc to allow comparisons
with more nearby cluster samples. Images were normalized by
0.3--7.0~keV exposure maps and the average background values listed in
Table~\ref{tab:obs} were subtracted off. Errors were determined
through 1000 Monte Carlo simulations whereby we added Poisson noise to
each cluster image and recalculated power ratios. As zero is a very
poor estimate of the Poisson noise, we instead adopted the average
background value per pixel (typically quite small) for each empty
pixel of a given observation.

We find that the power ratios from this sample occupy nearly identical
parameter space to those from the nearby sample of clusters from
\citet{Buote1996}. The only deviations are from the extremely compact
clusters A0115a, A2294, and RXJ0439.0$+$0520 which extend to very low
$P_{2}/P_{0}$ values. Overall, the majority of clusters in our sample
lie in the region of power ratio parameter space typical of relaxed,
symmetric clusters.

\begin{table*}
\centering
\begin{minipage}{114mm}
\caption{Power Ratios for the Cool Core Sample.}
\label{tab:props2}
\begin{tabular}{lcccc}
\hline
(1)            & (2)                 & (3)                 & (4)                 \\
Cluster        & $P_{2}/P_{0}$       & $P_{3}/P_{0}$       & $P_{4}/P_{0}$       \\
\hline
A68            & $(1.21^{+0.17}_{-0.14})\times10^{ -5}$ & $(1.02^{+0.80}_{-0.55})\times10^{ -7}$ & $(5.80^{+3.94}_{-2.99})\times10^{ -8}$ \\
A115a          & $(2.40^{+1.12}_{-0.83})\times10^{ -7}$ & $(1.86^{+0.48}_{-0.41})\times10^{ -7}$ & $(1.18^{+0.24}_{-0.21})\times10^{ -7}$ \\
A115b          & $(3.72^{+0.52}_{-0.51})\times10^{ -6}$ & $(3.87^{+0.91}_{-0.91})\times10^{ -7}$ & $(1.09^{+0.34}_{-0.30})\times10^{ -7}$ \\
A267           & $(7.23^{+1.17}_{-1.11})\times10^{ -6}$ & $(3.14^{+3.75}_{-2.22})\times10^{ -8}$ & $(2.25^{+2.09}_{-1.44})\times10^{ -8}$ \\
A520           & $(6.08^{+1.14}_{-1.04})\times10^{ -6}$ & $(2.88^{+1.47}_{-1.18})\times10^{ -7}$ & $(1.76^{+0.65}_{-0.56})\times10^{ -7}$ \\
A586           & $(8.18^{+2.25}_{-2.10})\times10^{ -7}$ & $(2.45^{+2.17}_{-1.39})\times10^{ -8}$ & $(1.56^{+1.18}_{-0.74})\times10^{ -8}$ \\
A665           & $(3.27^{+0.32}_{-0.27})\times10^{ -6}$ & $(1.64^{+0.35}_{-0.33})\times10^{ -7}$ & $(1.33^{+0.21}_{-0.20})\times10^{ -7}$ \\
A697           & $(6.55^{+0.71}_{-0.69})\times10^{ -6}$ & $(1.56^{+0.64}_{-0.51})\times10^{ -7}$ & $(1.47^{+1.12}_{-0.83})\times10^{ -8}$ \\
A750           & $(2.33^{+0.28}_{-0.25})\times10^{ -6}$ & $(4.25^{+2.03}_{-1.47})\times10^{ -8}$ & $(7.68^{+9.99}_{-5.63})\times10^{-10}$ \\
A773           & $(6.11^{+0.87}_{-0.83})\times10^{ -6}$ & $(2.94^{+3.19}_{-1.83})\times10^{ -8}$ & $(6.97^{+3.29}_{-2.78})\times10^{ -8}$ \\
A781           & $(4.67^{+1.99}_{-1.65})\times10^{ -6}$ & $(1.44^{+1.78}_{-1.01})\times10^{ -7}$ & $(4.18^{+2.14}_{-1.71})\times10^{ -7}$ \\
A963           & $(7.00^{+1.13}_{-0.93})\times10^{ -7}$ & $(5.11^{+1.36}_{-1.25})\times10^{ -8}$ & $(1.51^{+1.41}_{-0.87})\times10^{ -9}$ \\
A1204          & $(3.72^{+0.69}_{-0.59})\times10^{ -7}$ & $(1.18^{+0.61}_{-0.47})\times10^{ -8}$ & $(6.91^{+2.97}_{-2.64})\times10^{ -9}$ \\
A1423          & $(3.65^{+0.84}_{-0.75})\times10^{ -6}$ & $(3.59^{+3.93}_{-2.51})\times10^{ -8}$ & $(1.96^{+2.11}_{-1.31})\times10^{ -8}$ \\
A1682          & $(1.79^{+0.34}_{-0.30})\times10^{ -5}$ & $(1.36^{+0.51}_{-0.39})\times10^{ -6}$ & $(3.37^{+1.60}_{-1.24})\times10^{ -7}$ \\
A1758a         & $(4.20^{+0.16}_{-0.13})\times10^{ -5}$ & $(2.46^{+0.58}_{-0.50})\times10^{ -7}$ & $(4.66^{+0.56}_{-0.45})\times10^{ -7}$ \\
A1763          & $(9.74^{+0.98}_{-0.96})\times10^{ -6}$ & $(9.63^{+6.42}_{-4.36})\times10^{ -8}$ & $(3.67^{+2.25}_{-1.69})\times10^{ -8}$ \\
A1835          & $(6.26^{+0.51}_{-0.55})\times10^{ -7}$ & $(2.38^{+0.56}_{-0.51})\times10^{ -8}$ & $(6.72^{+5.69}_{-4.20})\times10^{-10}$ \\
A1914          & $(1.94^{+0.24}_{-0.22})\times10^{ -6}$ & $(1.17^{+0.29}_{-0.31})\times10^{ -7}$ & $(1.12^{+1.79}_{-0.77})\times10^{-10}$ \\
A2111          & $(1.45^{+0.25}_{-0.19})\times10^{ -5}$ & $(1.47^{+1.19}_{-0.79})\times10^{ -7}$ & $(1.04^{+0.68}_{-0.51})\times10^{ -7}$ \\
A2204          & $(5.32^{+0.48}_{-0.52})\times10^{ -7}$ & $(1.05^{+0.37}_{-0.30})\times10^{ -8}$ & $(7.64^{+1.87}_{-1.83})\times10^{ -9}$ \\
A2218          & $(2.65^{+0.20}_{-0.19})\times10^{ -6}$ & $(1.15^{+0.24}_{-0.19})\times10^{ -7}$ & $(1.98^{+1.73}_{-1.14})\times10^{ -9}$ \\
A2219          & $(2.08^{+0.06}_{-0.05})\times10^{ -5}$ & $(1.61^{+0.24}_{-0.22})\times10^{ -7}$ & $(1.39^{+0.14}_{-0.13})\times10^{ -7}$ \\
A2259          & $(4.08^{+0.70}_{-0.65})\times10^{ -6}$ & $(1.28^{+0.58}_{-0.49})\times10^{ -7}$ & $(1.04^{+1.15}_{-0.67})\times10^{ -8}$ \\
A2261          & $(7.42^{+2.42}_{-1.94})\times10^{ -7}$ & $(4.14^{+2.76}_{-2.02})\times10^{ -8}$ & $(1.19^{+1.04}_{-0.68})\times10^{ -8}$ \\
A2294          & $(4.96^{+7.17}_{-3.63})\times10^{ -8}$ & $(6.76^{+1.08}_{-0.51})\times10^{ -9}$ & $(8.12^{+1.38}_{-0.57})\times10^{-10}$ \\
A2390          & $(7.12^{+0.39}_{-0.37})\times10^{ -6}$ & $(2.15^{+1.23}_{-0.91})\times10^{ -8}$ & $(3.09^{+0.83}_{-0.72})\times10^{ -8}$ \\
Hercules A     & $(1.18^{+0.14}_{-0.14})\times10^{ -6}$ & $(6.83^{+6.67}_{-4.49})\times10^{ -9}$ & $(4.91^{+6.63}_{-3.46})\times10^{-10}$ \\
RXJ0439.0+0520 & $(8.61^{+7.93}_{-5.52})\times10^{ -8}$ & $(7.13^{+9.24}_{-5.19})\times10^{ -9}$ & $(1.13^{+1.73}_{-0.82})\times10^{ -9}$ \\
RXJ0439.0+0715 & $(3.32^{+0.41}_{-0.41})\times10^{ -6}$ & $(1.18^{+0.42}_{-0.35})\times10^{ -7}$ & $(1.25^{+0.83}_{-0.66})\times10^{ -8}$ \\
RXJ1532.9+3021 & $(6.07^{+1.18}_{-1.07})\times10^{ -7}$ & $(1.06^{+1.74}_{-0.75})\times10^{ -9}$ & $(4.28^{+5.31}_{-3.17})\times10^{-10}$ \\
RXJ1720.1+2638 & $(9.77^{+0.62}_{-0.61})\times10^{ -7}$ & $(5.89^{+2.61}_{-2.11})\times10^{ -9}$ & $(4.57^{+1.43}_{-1.21})\times10^{ -9}$ \\
RXJ2129.6+0005 & $(1.87^{+0.39}_{-0.34})\times10^{ -6}$ & $(4.61^{+8.10}_{-3.29})\times10^{ -9}$ & $(2.66^{+1.52}_{-1.17})\times10^{ -8}$ \\
Z1953          & $(4.02^{+0.79}_{-0.65})\times10^{ -6}$ & $(1.64^{+0.91}_{-0.66})\times10^{ -7}$ & $(1.77^{+1.49}_{-1.11})\times10^{ -8}$ \\
Z2701          & $(1.07^{+0.08}_{-0.09})\times10^{ -6}$ & $(8.51^{+10.6}_{-0.59})\times10^{-10}$ & $(4.13^{+4.12}_{-2.57})\times10^{-10}$ \\
Z3146          & $(6.68^{+0.71}_{-0.69})\times10^{ -7}$ & $(6.68^{+3.75}_{-2.86})\times10^{ -9}$ & $(2.05^{+1.46}_{-0.91})\times10^{ -9}$ \\
Z5247          & $(7.01^{+0.93}_{-0.92})\times10^{ -5}$ & $(1.94^{+2.28}_{-1.27})\times10^{ -7}$ & $(1.49^{+0.51}_{-0.41})\times10^{ -6}$ \\
Z7160          & $(3.99^{+1.36}_{-1.25})\times10^{ -7}$ & $(3.89^{+6.16}_{-2.70})\times10^{ -9}$ & $(3.33^{+3.74}_{-2.25})\times10^{ -9}$ \\
\hline
\end{tabular}

\smallskip 
{\bf Col. 1:} Cluster name. Alternate names are given in brackets.
{\bf Cols. 2--4:} Power ratios as determined in
Sect.~\ref{sec:power} for an aperture of 0.5~Mpc.
\end{minipage}
\end{table*}
 
\subsection{Spectral Analysis}\label{sec:spectra}

Spectral analysis of the clusters was performed following two methods.
For simplicity, X-ray spectral analysis in both cases was always
performed only on the first entry in Table~\ref{tab:obs} (typically
the longest observation) and data were masked out beyond intrinsic
radii of 0.5~Mpc to simplify and speed up analyses.

In the first method, we divided the data into several regions using a
``contour binning'' method \citetext{\citealp{Sanders2005a}; 2005b, in
preparation} such that the signal-to-noise was 20--40 per region
depending on the quality of the data.  The algorithm defines regions
with a signal-to-noise greater than a threshold by growing bins in the
direction on a smoothed map which has a value closest to the mean
value of those pixels already binned. This technique defines bins
which are matched to the surface brightness distribution of the
object.  X-ray spectra were extracted for each region and fit in XSPEC
with a single-temperature MEKAL model absorbed by a PHABS model. The
temperature in each region was allowed to vary between 0.1--20~keV,
while the absorption was fixed at the Galactic value along the line of
sight \citep[taken from][]{Ebeling1998}, the abundance was fixed to 30
per cent of solar,\footnote{Letting the abundance vary makes little
  difference to the measured temperatures.} and the redshift was set
to its appropriate value for each cluster. The results of this
spectral analysis were combined together to form the projected
temperature maps shown in Figure~\ref{fig:images}. The projected
temperature maps complement the colour images in terms of identifying
clusters with strong cool cores, but are particularly useful for
revealing weak cool cores and further substructure not readily
apparent from the images alone.

In the second method, we split the data into no more than 10
concentric circular annuli again such that the signal-to-noise was at
least 20 per region, but often much higher depending on the quality of
the data.\footnote{Most of the clusters in our sample are in fact
elliptical rather than spherical. However, for simplicity we have
derived profiles using circular rather than elliptical annuli since we
do not know whether the clusters are prolate or oblate. To determine
the degree of error this might introduce, we fitted elliptical annuli
to a few of the most eccentric clusters in our sample. We found
virtually no change in the derived profiles; the circular and
elliptical curves were consistent with each other to at least a
$\approx1\sigma$ confidence level with no apparent systematics.} X-ray
spectra were extracted for each region and these regions were fit
simultaneously in XSPEC with a PROJCT model, adopting absorbed (PHABS)
single-temperature MEKAL models for each annular bin (i.e., a model
identical to that used in the first method). Using these annular
spectral fits, we derived deprojected radial temperature, density,
cooling time, and mass deposition rate profiles.

\section{Discussion}\label{sec:discussion}

As found in Sections~\ref{sec:prep} and \ref{sec:power}, the majority
of the clusters have relatively regular morphologies and generally
appear to be relaxed.  From a visual inspection of the 38 BCS
clusters, we find five that are obviously double-peaked and four more that
have disturbed morphologies indicative of merging (see
Table~\ref{tab:props1}. The temperature maps in a handful of clusters
hint at mild additional substructure within 0.5~Mpc.  We note that our
assumption of spherical symmetry does little to account for clusters
with double-peaked or merging morphologies, which appear to make up
$\approx25$ per cent of the sample. In these cases, our radial
profiles simply serve as a proxy and are likely to overestimate the
real temperature and underestimate the real central cooling time. Thus
we caveat that cooling times for disturbed clusters should be regarded
as likely lower limits. We still provide this analysis since it may be
useful to have some point of reference for comparison to
lower-resolution nearby clusters observed with, e.g., {\it ROSAT} or
high-redshift clusters observed with {\it XMM-Newton} or {\it
Chandra}, where statistics are generally poorer and mergers less
obvious.

\begin{figure}
\vspace{-0.1in}
\centerline{
\includegraphics[width=9.0cm]{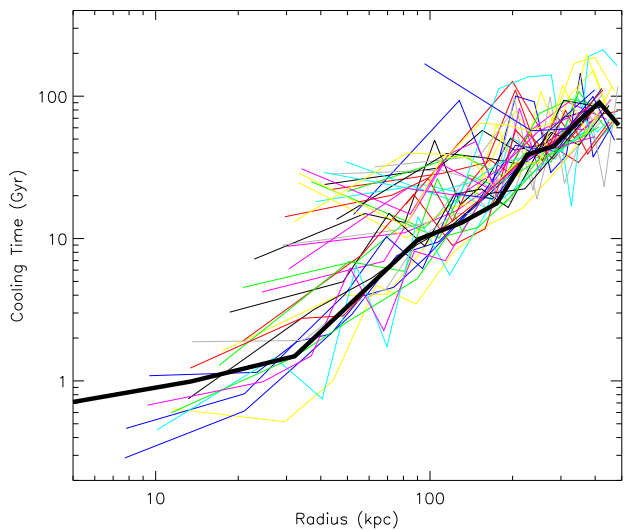}
}
\vspace{-0.1cm} \caption[fig4.ps]{Radial cooling time profiles for
  each of the 38 BCS clusters in the sample (denoted by the different
  grey/colour lines, for paper/online publishing respectively). There
  appears to be a lower bound to the cooling profiles.  The thick
  black line, which traces this lower bound, denotes the average
  cooling profile for the clusters with a central cooling time less
  than 2~Gyr.
  \label{fig:cooling_profiles}}
\vspace{-0.3cm}
\end{figure} 

\subsection{Cooling Time Profiles}\label{sec:cool_profs}

We have measured the cooling time profiles for all of the clusters in
our sample, as shown in Figure~\ref{fig:cooling_profiles}. While the
cooling times in the central bin differ by a few orders in magnitude,
we note that there appears to be a prominent, apparently universal
lower bound to all of the cooling profiles. The average cooling time
profile for the 13 clusters with a central cooling time less than
2~Gyr (thick black line) indicates that there is little difference in
the lower bounds for a set of clusters with widely ranging physical
characteristics (i.e., temperatures, densities, entropies).  While we
might expect some sort of boundary given the general nature of how gas
cools in clusters, it is physically unclear why a universal boundary
that takes the form $t_{c} \propto r^{1.3-1.5}$ (depending on whether
one imposes an inner cutoff) would exist. Similarly, \citet{Voigt2004}
find a very small dispersion in cooling time profiles among their
sample of 16 clusters with strong cool cores, with $t_{c} \propto
r^{1.3}$.

\begin{figure}
\vspace{-0.1in}
\centerline{
\includegraphics[width=9.0cm]{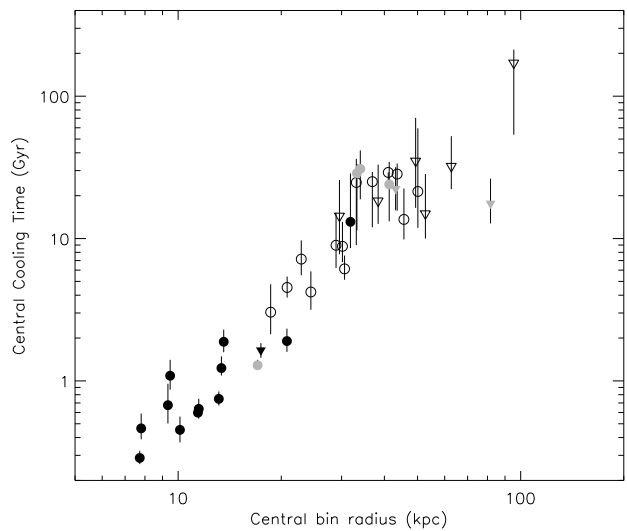}
}
\vspace{-0.1cm} \caption[fig5.ps]{Plot of the central cooling time for
  the inner radial bin for the 38 BCS clusters in the sample. 
  Filled black, empty, and filled grey symbols
  denote clusters with detected, undetected, and unobserved H$\alpha$
  line emission, respectively \citep{Crawford1999}. Circles denote
  morphologically regular clusters, while triangles denote
  double-peaked or disturbed clusters. There is a strong
  correspondence between low cooling time and detection of H$\alpha$
  line emission. The one exception is A2294. 
  \label{fig:central_cooling_time}}
\vspace{-0.3cm}
\end{figure} 

The cooling times for the innermost radial bin are shown in
Figure~\ref{fig:central_cooling_time} for each cluster. Unfortunately
in physical units, the innermost radial bin is both a function of the
cluster distance and the overall signal-to-noise of the observation
(which in turn depends on whether a cluster exhibits a cool core);
hence the apparent correlation in
Figure~\ref{fig:central_cooling_time}.  Importantly, the spatial
resolution of {\it Chandra} is a factor of several smaller than the
typical innermost radius adopted (i.e., 0\farcs5 corresponds to
1--3~kpc), so any failure to detect a significant cool core in the BCS
sample is not due to any limitation on spatial resolution. Also note
that because we use the cluster centroid rather than the surface
brightness peak, the cooling times inside a few tens of kpc could
be biased upward by at most a factor of $\sim$2--3.\footnote{The peak
and centroid surface brightnesses ($\propto n^{2}$) and cluster
temperatures never differ by more than a factor of 2 each. Since
$t_{c} \propto T^{0.5} n^{-1}$, this means that $t_{c}$ is unlikely to
change dramatically based on our adopted position.  Significant
differences between the peak and centroid positions only occur in
highly disturbed or double-peaked clusters and comprise at worst
$\la20$ per cent of our sample (see Table~\ref{tab:props1}).}
Alternatively, the central cooling times should be considered upper
limits because the strength of cooling depends strongly on the radius
at which it is measured, and thus on the signal-to-noise of the
observation.

\begin{figure*}
\vspace{-0.1in}
\centerline{
\includegraphics[width=6.5cm]{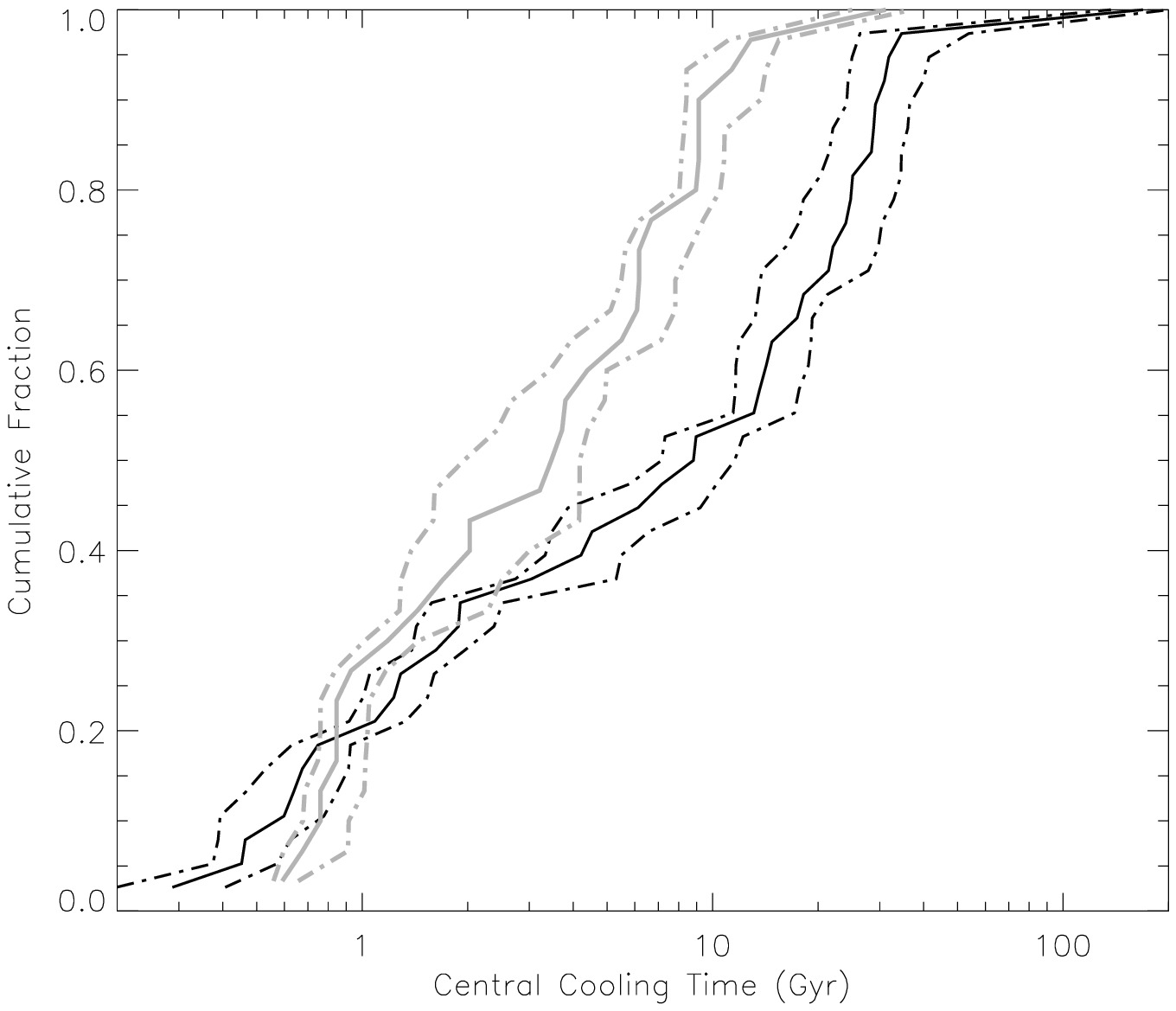}
\hglue-0.2in{\includegraphics[width=6.5cm]{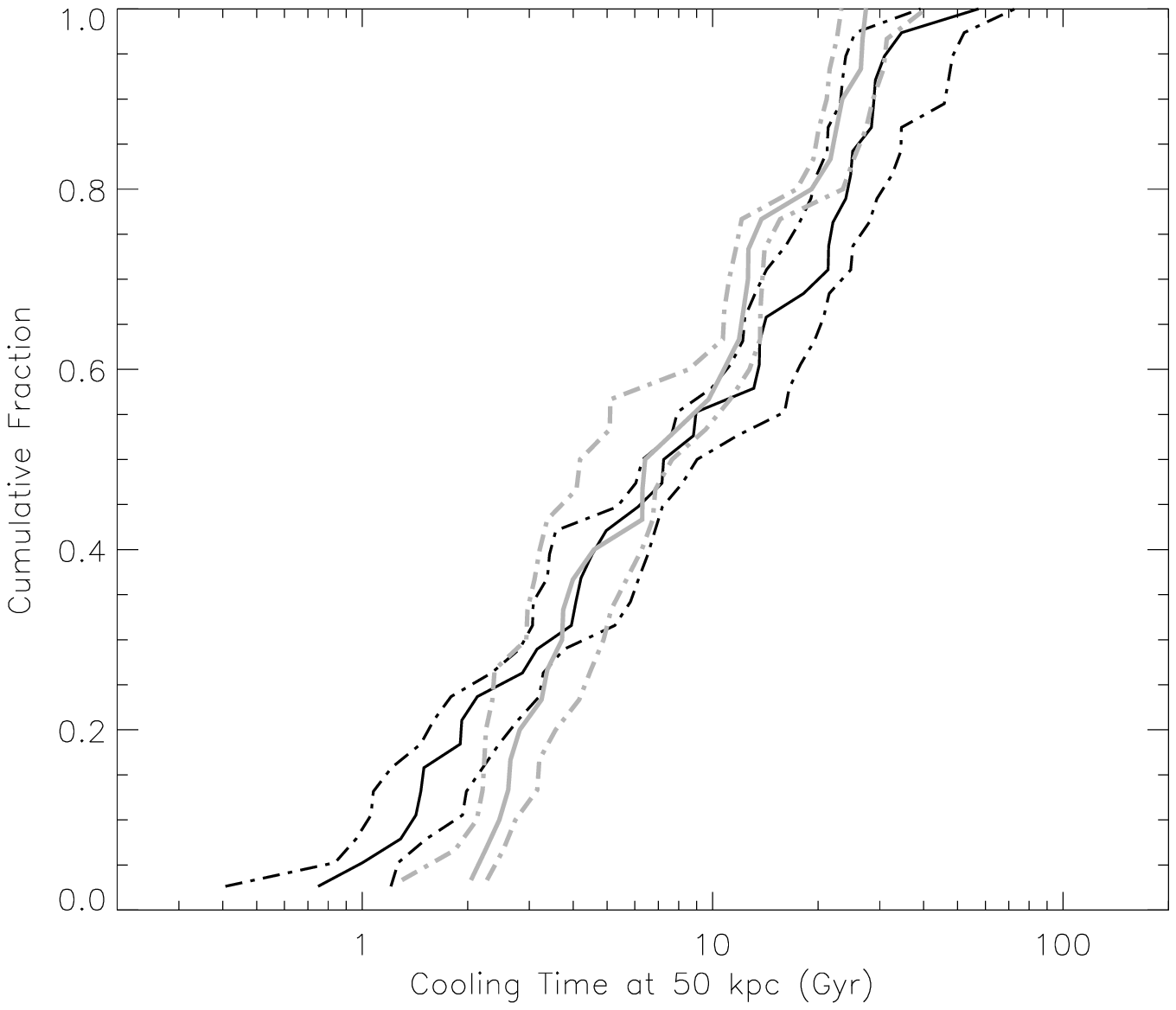}}
\hglue-0.2in{\includegraphics[width=6.5cm]{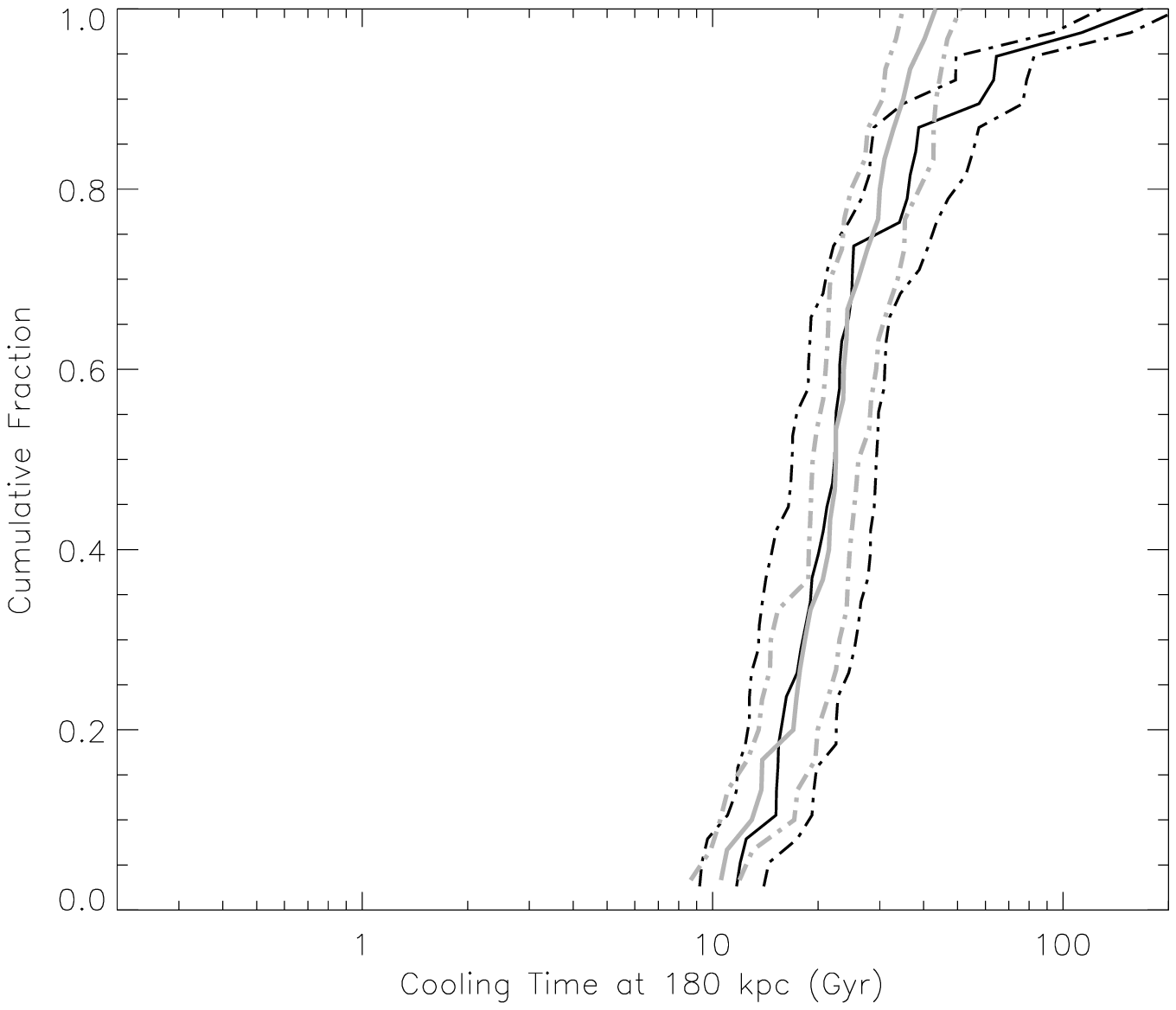}}
}
\vspace{-0.1cm} \caption[fig6.ps]{Comparison of the cumulative
  fraction of cooling times ({\it left:} central bin; {\it middle:} at
  50~kpc; {\it right:} at 180~kpc) for the 38 BCS clusters in the
  sample (black) and the B55 sample of \citet[grey;][]{Peres1998}. The
  dash-dotted lines denote the 1~$\sigma$ error bars. The large apparent
  difference seen in the cumulative fraction measured for the central
  bin disappears when comparisons are made at fixed intrinsic radii.
  \label{fig:cooling_time_comp}}
\vspace{-0.3cm}
\end{figure*}

From our determination of cooling times for the innermost radial bin,
we find that at least 20 of 38 (53 per cent) BCS clusters in the {\it
  Chandra} sample have central cooling times $<10$~Gyr (mild cool
cores) and at least 13 of 38 (34 per cent) have $<2$~Gyr (strong cool
cores). To test for any evolution, we compare the cumulative cooling
time fractions from the BCS sample to those found for the nearby B55
sample of \citet{Peres1998}, who provide cooling times for both the
innermost radial bin and 180~kpc (originally 250~kpc in their
cosmology).  At first glance, the fraction of cool core clusters in
our sample as measured from the innermost radial bin would appear to
be markedly different from those in the B55 sample (see left plot in
Figure~\ref{fig:cooling_time_comp}), implying substantial evolution in
the cores of X-ray bright clusters out to $z\sim0.4$. However, we must
exercise caution here, since a direct comparison between the
two samples in this manner is problematic.  Importantly, the innermost
radial bins span a large range in both samples and are significantly
smaller in the B55; thus comparing these quantities is not likely to
provide a useful yardstick for cooling.  Additionally, we note that
\citet{Peres1998} measured their cooling times using the surface
brightness peak rather than the centroid, so there may be a slight
systematic difference between the calculated cooling times for a small
fraction of clusters. A final consideration is that the {\it ROSAT}
observations of the fainter or more distant clusters in the B55 sample
may fail to adequately resolve potential cool cores due to limited
spatial resolution.

Given the above problems, we instead choose to compare cooling times
for the two samples at a fixed intrinsic radius where cooling is still
important. Since a large fraction of both samples have innermost
radial bins close to or smaller than 50~kpc, this value seems like a
sensible radius to use. At 50~kpc, we find that at least 21 of 38 (55
per cent) BCS clusters in the {\it Chandra} sample have central
cooling times $<10$~Gyr (mild cool cores) and at least 8 of 38 (21 per
cent) have $<2$~Gyr (strong cool cores).  Because the B55 sample lacks
published cooling time profiles, we must interpolate cooling times at
50~kpc.  For this purpose, we adopt as an interpolation template the
average cooling profile found in Figure~\ref{fig:cooling_profiles} for
clusters with $t_{c}<2$~Gyr.  Comparisons at 50~kpc (middle), as well
as at 180~kpc (right), are shown in
Figure~\ref{fig:cooling_time_comp}. The apparent differences in the
cooling time fractions measured using the innermost radial bin have
now all but vanished, demonstrating that there is no detectable
evolution in the cool core fraction of clusters out to $z\sim0.4$.
Note that the small discrepancies for short cooling times may be due
to our assumed interpolation template. Also note that the fraction of
cool cores in either sample has no obvious X-ray luminosity
dependence.

Let us consider the evolution of a cluster core hosting a cooling
flow. First assume that it is isolated with no heating, so that
radiative cooling proceeds to drop the gas temperature. Approximating
the situation locally as a constant pressure flow, then the cooling
time at 50~kpc will decrease by approximately the mean time difference
between the samples. The mean redshift of the B55 sample is 0.056 and
of the 38 BCS clusters is 0.22, so this difference in our adopted
cosmology is about 1.8~Gyr. It is clear from
Fig.~\ref{fig:cooling_time_comp} that this is not occurring since at
the 50 per cent level the curves are consistent with each other.
Second, we consider the effect of the continual accretion of
subclusters by a cluster \citep[e.g.,][]{Rowley2004}, so that the cool
central region undergo adiabatic compression which, for bremsstrahlung
cooling appropriate here, varies as $T^{-1}$. This again reduces as
time proceeds. Weak to moderate shocks also reduce the cooling time.
What is needed is a non-gravitational heat source which offsets the
effects of radiative cooling. The similarity between the cumulative
curves at 50~kpc for the B55 and BCS samples demonstrates that the
heating invoked to balance radiative cooling required to explain the
detailed temperature profiles of the clusters must have been in place
before the redshift of the BCS.  More distant samples are required to
determine just when the heating/cooling balance was established.

\subsection{Ubiquity of a Cooling ``Floor''}\label{sec:cool_drop}

\begin{figure*}
\vspace{-0.1in}
\centerline{
\includegraphics[width=8.0cm]{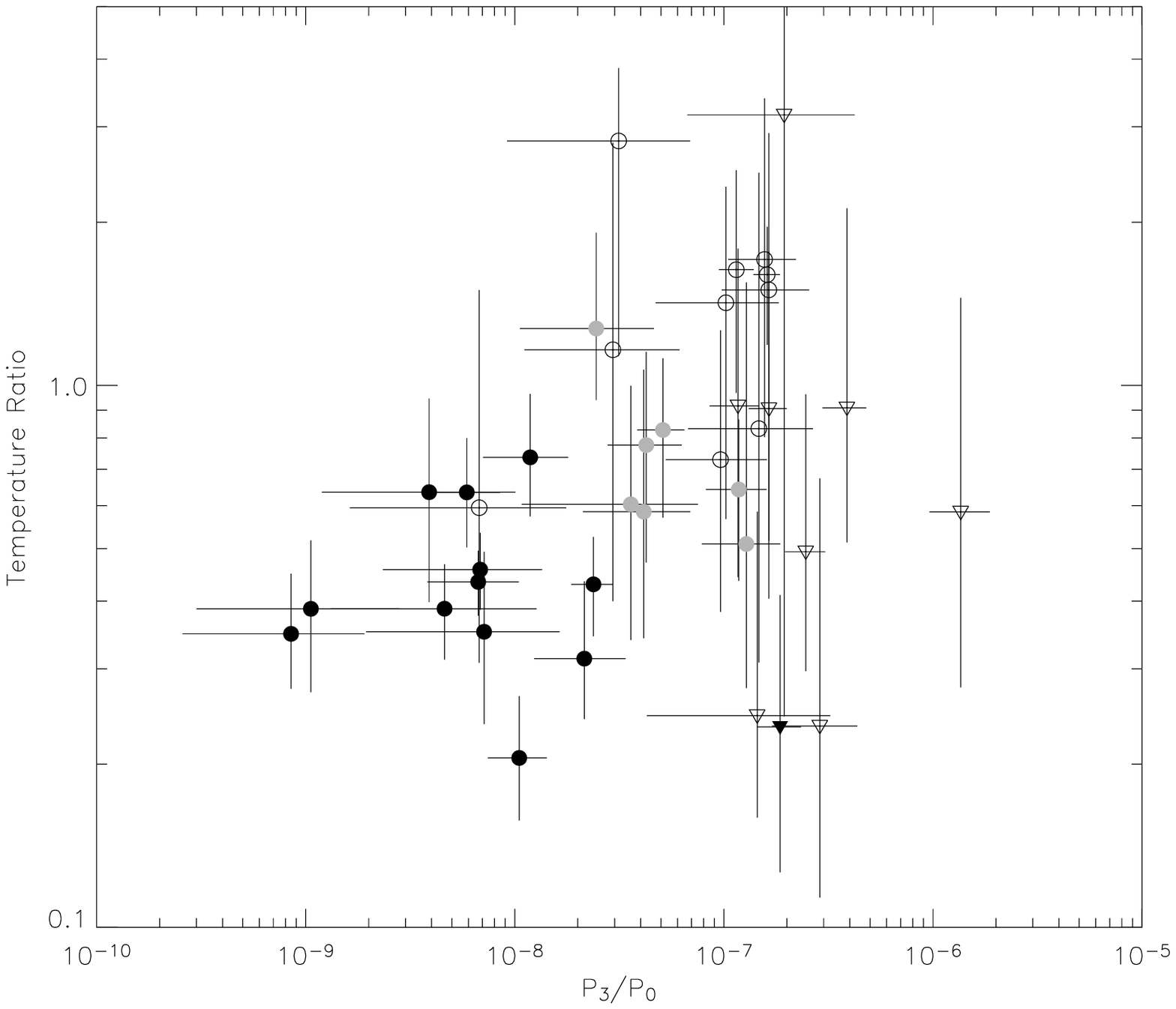}
\includegraphics[width=8.0cm]{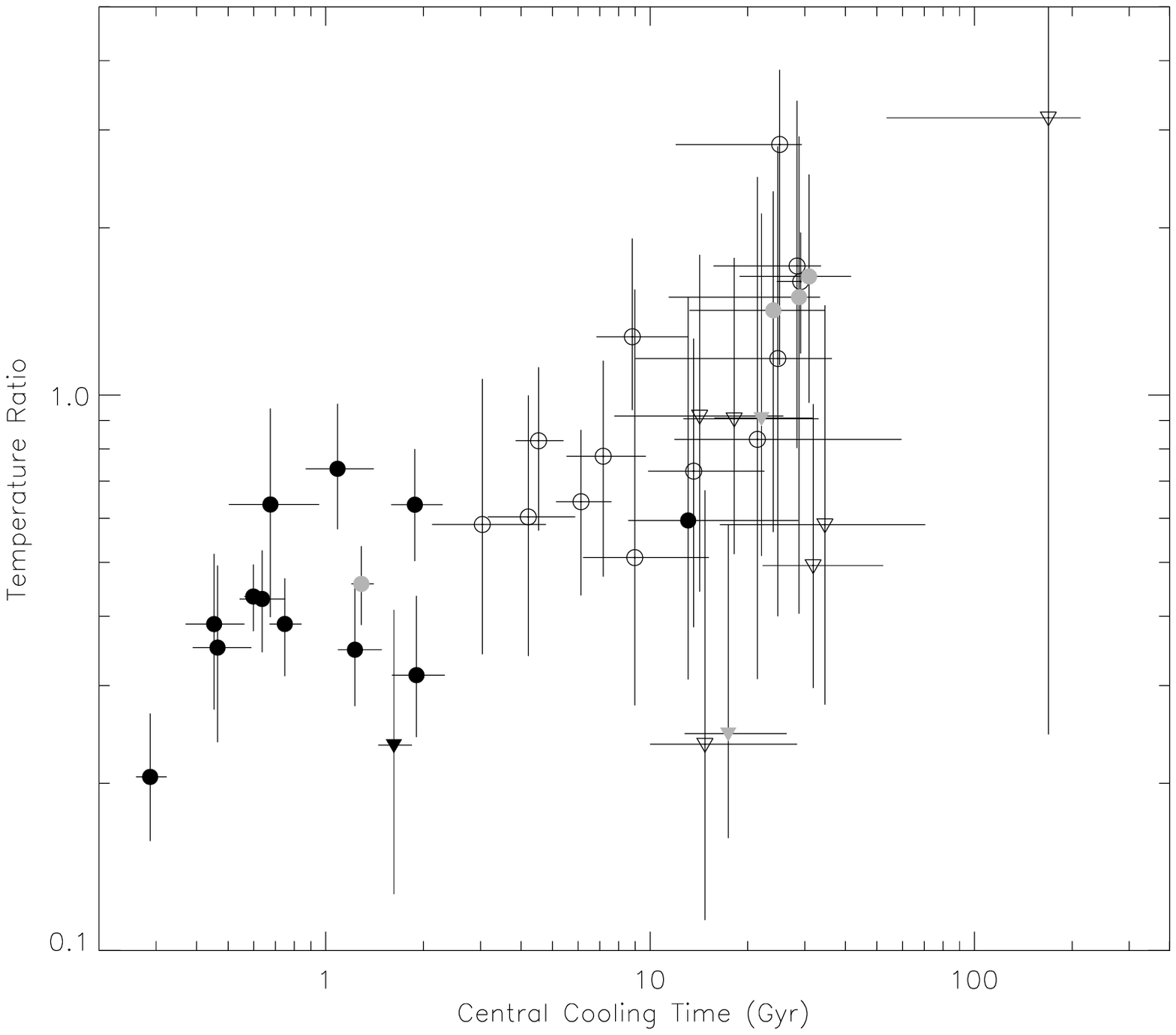}
}
\vspace{-0.1cm} \caption[fig7.ps]{Plot of the ratio between the
  central bin temperature and the average temperature outside of
  180~kpc versus 0.5~Mpc power ratios ({\it left}) and central cooling
  times ({\it right}) for the 38 BCS clusters in the sample. {\it
    Left}: Filled black, grey, and empty symbols denote clusters with
  strong cool cores ($t_{c}<2$~Gyr), weak cool cores
  ($t_{c}=2$--10~Gyr), and no obvious cool cores ($t_{c}>10$~Gyr),
  respectively. {\it Right}: Filled black, empty, and filled grey
  symbols denote clusters with detected, undetected, and unobserved
  H$\alpha$ line emission, respectively \citep{Crawford1999}. Circles
  denote morphologically regular clusters, while triangles denote
  double-peaked or disturbed clusters. \label{fig:cool_floor}}
\vspace{-0.3cm}
\end{figure*}

As noted in $\S$~\ref{sec:intro}, the central temperatures in clusters
are found to generally drop by less than a factor of $\sim$3--4 with
very little cooler gas \citep[e.g.,][]{Peterson2001, Tamura2001,
  Peterson2003, Kaastra2004}.  This cooling ``floor'' is thought to be
due to some form of heating (e.g., from an active galactic nucleus,
conduction).  Figure~\ref{fig:cool_floor} compares the temperature
drop --- taken to be ratio between the central bin temperature and the
average temperature outside of 180~kpc --- to both the 0.5~Mpc power
ratios and central cooling times ({\it right}) for the 38 BCS clusters
in the sample. A similar cooling floor (i.e., a temperature drop of no
more than $\approx3$--4) exists for our cluster sample, and does not
appear to correlate with either the strength of the cool core or its
general morphology. Interestingly, three of the four lowest
temperature drop clusters are merging clusters; inspection of the
temperature maps for these clusters indicates that their large
temperature ratios are in part likely due to shocks in the outer
regions of those clusters. In general, we find that the disturbed
clusters in our sample, which all have high $P_{3}/P_{0}$ values, span
the entire range of observed temperature ratios, presumably due to the
varied merger histories and ages of these clusters.

\subsection{Cooling Time vs. H$\alpha$}\label{sec:cool_halpha}

A significant fraction of clusters with cool cores have also been
shown to exhibit signs of filamentary H$\alpha$ line emission. Since
nearly all of the clusters in our sample have H$\alpha$ observations
\citep{Crawford1999}, we can examine how useful H$\alpha$ emission is
as an indicator of cool cores and cluster morphological properties.
Figure~\ref{fig:central_cooling_time} demonstrates that 13 of the 38
clusters in the sample show evidence for H$\alpha$ line emission and
all but one of these are among the clusters with the very shortest
central cooling times.\footnote{A2294 is the one outlier. This source
  has a strong detection (C. Crawford, private communication),
  although we are unable to rule out the possibility of contamination
  from nearby galaxies.} This result strengthens the correlation found
by \citet{Peres1998} for relatively nearby clusters.  Interestingly,
there are another three clusters among the 18 not observed by {\it
  Chandra} that also show evidence for H$\alpha$ line emission. If the
H$\alpha$-detected fraction traces the overall cluster population,
this would suggest that $\sim$15--20 per cent of the unobserved
sources are likely to have strong cool cores and that we might expect
a similar overall cool core distribution (i.e., both strong and
moderate) among the unobserved clusters as found for the sample
sources.

The observed H$\alpha$ and star formation rates requires the mass of
H$\alpha$-emitting gas typically to be only a few to a few tens of
$M_{\odot}$~yr$^{-1}$, whereas classical cooling flows in the observed
clusters would require many hundreds of $M_{\odot}$~yr$^{-1}$. Thus
the presence of H$\alpha$ could imply continuous cooling below the
``floor'' at a rate of up to a few tens of $M_{\odot}$~yr$^{-1}$. Or
alternatively, it could imply the presence of conditions necessary for
long-lived H$\alpha$.

\subsection{Comparison to Power Ratios}\label{sec:cool_power}

\begin{figure}
\vspace{-0.1in}
\centerline{
\includegraphics[width=8.0cm]{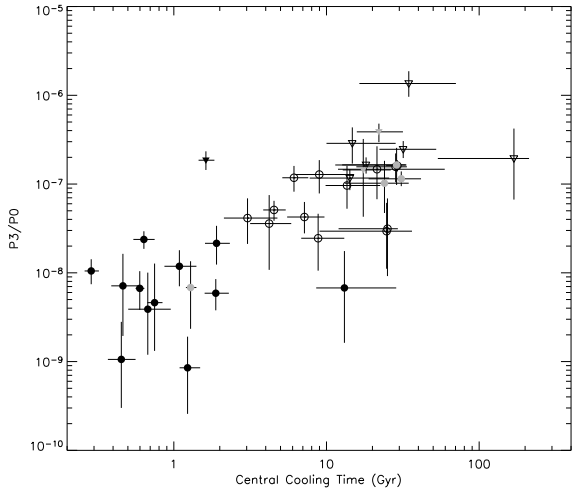}
}
\vspace{-0.1cm} \caption[fig8.ps]{Plot of the $P_{3}/P_{0}$ power
  ratios and central cooling times for the 38 BCS clusters in the
  sample. Filled black, empty, and filled grey symbols denote clusters
  with detected, undetected, and unobserved H$\alpha$ line emission,
  respectively \citep{Crawford1999}. Circles denote morphologically
  regular clusters, while triangles denote double-peaked or disturbed
  clusters. \label{fig:tc_p3p0}}
\vspace{-0.3cm}
\end{figure}

A large number of higher redshift clusters have now been found. As
these sources are less likely to have the signal-to-noise or spatial
resolution to unequivocally detect any potential cool cores, it is
instructive to compare our cool core estimates and H$\alpha$
detections for the BCS clusters with their power ratios (which can be
measured more reliably under the above conditions). Indeed,
Figure~\ref{fig:power2} demonstrates that the power ratios provide a
quantitive assessment of basic morphological trends and can be used as
a proxy for central cooling times. Figure~\ref{fig:tc_p3p0} bears out
this relationship better, showing the strongest correlation between
$P_{3}/P_{0}$ versus $t_{\rm cool}$.  As expected
\citep[e.g.,][]{Buote1996}, the more symmetric, compact clusters
generally have stronger cool cores and a higher likelihood of
H$\alpha$ line detections.

%
%
%
%
%
%
%
%

\section{Conclusions}\label{sec:conclusions}

Using a sample of 38 luminous BCS clusters at $z=0.15$--0.4, we find
that cool cores still appear to be common at moderate redshift. At
50~kpc, at least 55 per cent of the clusters in our sample possess
mild cool cores ($t_{\rm cool} < 10$~Gyr) and within the central bin
at least 34 per cent possess strong cool cores ($t_{\rm cool} <
2$~Gyr). We find that a cooling floor exists for our moderate redshift
sample, similar to that found for many nearby clusters, such that the
ratio of central to outer temperatures rarely increases above a factor
of $\approx$~3--4. Moreover, comparing the central cooling times to
catalogues of central H$\alpha$ emission in BCS clusters, we find a
strong correspondence between the detection of H$\alpha$ and short
central cooling times. We also find a correlation between the central
cooling time and cluster power ratios, indicating that crude
morphological measures are a proxy for more rigorous analysis in the
face of limited signal-to-noise data.

Comparing our moderate redshift BCS sample to the local B55 sample
($z<0.15$), we find no evidence for any significant evolution in the
cumulative cooling time fractions at both 50~kpc and 180~kpc. This
suggests that cool cores are already well-developed even at moderate
redshifts and, as a consequence, are quite likely to be robust against a
wide variety of merger scenarios. It also implies that any additional
cooling our sample of clusters underwent over the last $\sim$2--3 Gyr
must be balanced by some form of heating. Thus, balanced heating and
cooling mechanisms are likely to have already stabilised or been
frozen in by this epoch.

Looking toward the future, it will be instructive to analyse {\it
  Chandra} and {\it XMM-Newton} observations for a sample of local
clusters (such as the B55 sample) in a manner similar to that done
here for the BCS sample, so we can perform a more complete and
rigorous comparison of the two samples. Further work is also needed to
understand when the onset of cool cores occurs.  The high fraction of
cooling core clusters observed here needs to be tied to the relative
dearth of such clusters at the highest redshifts ($z\sim1$).  Another
issue is to determine when and how the heating process, which leads to
the nearly ubiquitous temperature plateau of a factor of 3--4, occurs.
This heating process appears to already be in place for the clusters
in our sample, so again we must look to higher redshift samples for
clues.

\section{Acknowledgements}

FEB and RMJ acknowledge support from PPARC. ACF and SWA thank the
Royal Society for support.


\begin{thebibliography}{}

\bibitem[Allen \& Fabian(1997)]{Allen1997} Allen, S.~W.~\& 
Fabian, A.~C.\ 1997, \mnras, 286, 583 

\bibitem[Allen et al.(2001)]{Allen2001} Allen, S.~W., Fabian, 
A.~C., Johnstone, R.~M., Arnaud, K.~A., \& Nulsen, P.~E.~J.\ 2001, \mnras, 
322, 589 

\bibitem[Buote \& Tsai(1996)]{Buote1996} Buote, D.~A.~\& Tsai, 
J.~C.\ 1996, \apj, 458, 27 

\bibitem[Crawford et al.(1999)]{Crawford1999} Crawford, C.~S., 
Allen, S.~W., Ebeling, H., Edge, A.~C., \& Fabian, A.~C.\ 1999, \mnras, 
306, 857 

\bibitem[Ebeling et al.(1998)]{Ebeling1998} Ebeling, H., Edge, 
A.~C., Bohringer, H., Allen, S.~W., Crawford, C.~S., Fabian, A.~C., Voges, 
W., \& Huchra, J.~P.\ 1998, \mnras, 301, 881 

\bibitem[Ebeling et al.(2000)]{Ebeling2000} Ebeling, H., Edge, 
A.~C., Allen, S.~W., Crawford, C.~S., Fabian, A.~C., \& Huchra, J.~P.\ 
2000, \mnras, 318, 333 

\bibitem[Edge, Stewart, \& Fabian(1992)]{Edge1992} Edge, A.~C., 
Stewart, G.~C., \& Fabian, A.~C.\ 1992, \mnras, 258, 177 

\bibitem[Fabian(1994)]{Fabian1994} Fabian, A.~C.\ 1994, \araa, 32, 
277 

\bibitem[G{\' o}mez, Loken, Roettiger, \& Burns(2002)]{Gomez2002} 
G{\' o}mez, P.~L., Loken, C., Roettiger, K., \& Burns, J.~O.\ 2002, \apj, 
569, 122 

\bibitem[Jerius et al.(2000)]{Jerius2000} Jerius, D., Donnelly, 
R.~H., Tibbetts, M.~S., Edgar, R.~J., Gaetz, T.~J., Schwartz, D.~A., Van 
Speybroeck, L.~P., \& Zhao, P.\ 2000, Proc. SPIE Vol. 4012, 17 

\bibitem[Kaastra et al.(2004)]{Kaastra2004} Kaastra, J.~S., et al.\ 
2004, \aap, 413, 415 

\bibitem[Lupton et al.(2004)]{Lupton2004} Lupton, R., Blanton, 
M.~R., Fekete, G., Hogg, D.~W., O'Mullane, W., Szalay, A., \& Wherry, N.\ 
2004, \pasp, 116, 133 

\bibitem[McGlynn \& Fabian(1984)]{McGlynn1984} McGlynn, T.~A.~\& 
Fabian, A.~C.\ 1984, \mnras, 208, 709 

\bibitem[Peres et al.(1998)]{Peres1998} Peres, C.~B., Fabian, 
A.~C., Edge, A.~C., Allen, S.~W., Johnstone, R.~M., \& White, D.~A.\ 1998, 
\mnras, 298, 416 

\bibitem[Peterson et al.(2001)]{Peterson2001} Peterson, J.~R., et 
al.\ 2001, \aap, 365, L104 

\bibitem[Peterson et al.(2003)]{Peterson2003} Peterson, J.~R., Kahn, 
S.~M., Paerels, F.~B.~S., Kaastra, J.~S., Tamura, T., Bleeker, J.~A.~M., 
Ferrigno, C., \& Jernigan, J.~G.\ 2003, \apj, 590, 207 

\bibitem[Ritchie \& Thomas(2002)]{Ritchie2002} Ritchie, B.~W.~\& 
Thomas, P.~A.\ 2002, \mnras, 329, 675 

\bibitem[Rowley, Thomas, \& Kay(2004)]{Rowley2004} Rowley, D.~R., 
Thomas, P.~A., \& Kay, S.~T.\ 2004, \mnras, 352, 508 

\bibitem[Sanders et al.(2005)]{Sanders2005a} Sanders, J.~S., Fabian, 
A.~C., \& Taylor, G.~B.\ 2005, \mnras, 356, 1022 

\bibitem[Tamura et al.(2001)]{Tamura2001} Tamura, T., et al.\ 
2001, \aap, 365, L87 

\bibitem[Townsley et~al.(2002)]{Townsley2002}
{Townsley}, L.~K., {Broos}, P.~S., {Nousek}, J.~A., \& {Garmire}, G.~P. 2002,
  Nuclear Instruments and Methods in Physics Research A, 486, 751

\bibitem[Voigt \& Fabian(2004)]{Voigt2004} Voigt, L.~M.~\& 
Fabian, A.~C.\ 2004, \mnras, 347, 1130 

\end{thebibliography}
\end{document}